\documentclass{aastex62}

\usepackage{epsfig}
\usepackage{color}
\usepackage{amsmath} 
\usepackage{amssymb} 
\usepackage{bm} 
\usepackage{enumerate}
\usepackage{float}
\usepackage{dirtytalk}

\newcommand{\p}{\partial}

\graphicspath{{./}{figures/}}

\received{\_\_\_\_  
}
\revised{\_\_\_\_ 
}
\accepted{\_\_\_\_ 
}
\submitjournal{ApJ}

\shorttitle{Hyper-Resistive Model of UHECR Acceleration by Magnetically Collimated AGN Jets }
\shortauthors{Fowler et al.}

\begin{document}

\title{Hyper-Resistive Model of Ultra High Energy Cosmic Ray Acceleration by Magnetically Collimated Jets Created by Active Galactic Nuclei
}

\correspondingauthor{Hui Li}
\email{hli@lanl.gov}

\author
{T. Kenneth Fowler}
\affil{University of California, Berkeley, CA, 94720 USA}

\author{Hui Li}
\affiliation{Los Alamos National Laboratory, Los Alamos, NM,  87545 USA}

\author{Richard Anantua}
\affil{University of California, Berkeley, CA, 94720 USA}
\affil{Harvard-Smithsonian Center for Astrophysics, Cambridge, MA, 02138 USA}

\begin{abstract}

This is the fourth in a series of companion papers showing that, when an efficient dynamo can be maintained 
by accretion disks around supermassive black holes in Active Galactic Nuclei (AGNs), it will lead to the formation of a powerful, magnetically-collimated helix that could explain both the observed jet/radiolobe structures on very large scales and ultimately the enormous power inferred from the observed ultra high energy cosmic rays (UHECRs) with energies $> 10^{19}\text{ eV}$. 
Many timescales are involved in this process. 
Our hyper-resistive magnetohydrodynamic (MHD) model provides a bridge between General 
Relativistic MHD simulations of dynamo formation, on the short accretion timescale, and observational evidence of magnetic collimation of large-scale jets on astrophysical timescales. 
Given the final magnetic structure, we apply hyper-resistive kinetic theory to show how instability causes slowly-evolving magnetically-collimated jets to become the most powerful relativistic accelerators in the Universe. 
The model yields nine observables in reasonable agreement with observations: the jet length, radiolobe radius and 
apparent opening angle as observed by synchrotron radiation; the synchrotron total power, synchrotron wavelengths 
and 
maximum electron energy (TeVs); and the maximum UHECR energy, the cosmic ray energy spectrum and the cosmic ray intensity on Earth.

\end{abstract}


\section{Introduction}\label{sec:intro}

The importance of understanding the origin of ultra high energy cosmic rays (UHECRs), with energies up to $\approx 10^{20}\text{ eV}$, has long been recognized \citep{cronin1999}.
Many explanations have been offered \citep{bierman1997}, none yet fully satisfactory \citep{blandford2019}.

This paper builds on \citet{colgate2004} in which it was hypothesized that UHECRs arise as electric currents in magnetic jets created by massive black holes inside Active Galactic Nuclei (AGNs), still a viable hypothesis \citep{blandford2017}.
Colgate and Li had already concluded that only AGNs with black hole masses $10^8$ to $10^9 M_{\odot}$ 
could account for the total energy in UHECRs. 
The discovery by \citet{balbus1998} that magneto-rotational (MRI) instability in rotating accretion disks around AGNs 
can create powerful dynamos strongly suggests that AGNs eject the magnetic jets observed by synchrotron radiation \citep{krolik1999}; and General Relativistic Magnetohydrodynamic (GRMHD) simulations have produced these dynamos \citep{mckinney2012}.
 Direct attempts to correlate UHECRs with known AGNs are limited by statistics but may yet pin down UHECR origins \citep{pierre2007,pierre2014,pierre2017}.
 
It has long been appreciated that AGN dynamos could produce the $10^{20}$ volts needed to accelerate UHECRs \citep{lovelace1976,lynden-bell2006}. 
The uncertainty has concerned what mechanism transfers dynamo voltage to ion acceleration. 
Unlike the transient mechanisms mentioned in \citet{colgate2004}, the mechanism proposed in this paper steadily accelerates ions ejected from the accretion disk over the full length of the jet, including the lobe regions. 
This acceleration mechanism is based on known plasma phenomena. 
Acceleration occurs in two stages. 
First is a weaker precursor stage due to current-driven magnetohydrodynamic (MHD) kink instability known to accelerate ions in the laboratory \citep{rusbridge1997}. 
This is followed by a powerful, purely kinetic stage based on a known ion cyclotron resonance instability that could be further explored in the laboratory, as discussed in \citet{fowler2016}.  

Confidence in our accelerator model depends on confidence in the jet structure undergirding this model. 
We begin this paper with a brief review of our model of the jet accelerator structure which is supported by the recent
new measurements from the RadioAstron mission \citep{giovannini2018}. 
This review appears in Section \ref{sec:towerstructure}, together with a discussion of 
MRI-driven dynamos in Appendix A. 
This is followed by a discussion of ion acceleration, in Section \ref{sec:ionacc} together with 
Appendix B; the predicted UHECR energy spectrum and cosmic ray intensity on Earth, 
in Section \ref{sec:CRs}; synchrotron radiation as a signature of our model, in Section 
\ref{sec:ele} and Appendix C; and a summary of this paper and all papers in this series, in Section \ref{sec:sum}. 

Our jet model is axisymmetric, though analogous WKB solutions would apply when jets are bent by encounters with the ambient \citep{begelman1984}. 
We use a stationary system of cylindrical coordinates $\{r,\phi, z\}$ in which the disk spins about a fixed $z$-axis with angular frequency $\bm{\Omega}$ pointing along the $+z$-direction in the inner region of the disk, giving positive toroidal magnetic field $B_{\phi}$ and negative $B_z$ in the same region. 
Except as noted, units are cgs, often introducing $c$, the speed of light.

\section{THE 
MAGNETIC ACCELERATOR STRUCTURE}
\label{sec:towerstructure}

We begin with a review of our disk/jet model in \citet{colgate2014}---hereafter Paper I--- and jet structure and stability in \citet{colgate2015}---hereafter Paper II. 

\subsection{Jets as Current Loops}
Together, Papers I and II provide the basis for 
the final quasi-static magnetic structure yielding steady-state acceleration of cosmic rays.

\begin{figure}\nonumber
\begin{align}
  \hspace{0cm} \includegraphics[height=350pt,trim = 6mm 1mm 0mm 1mm]{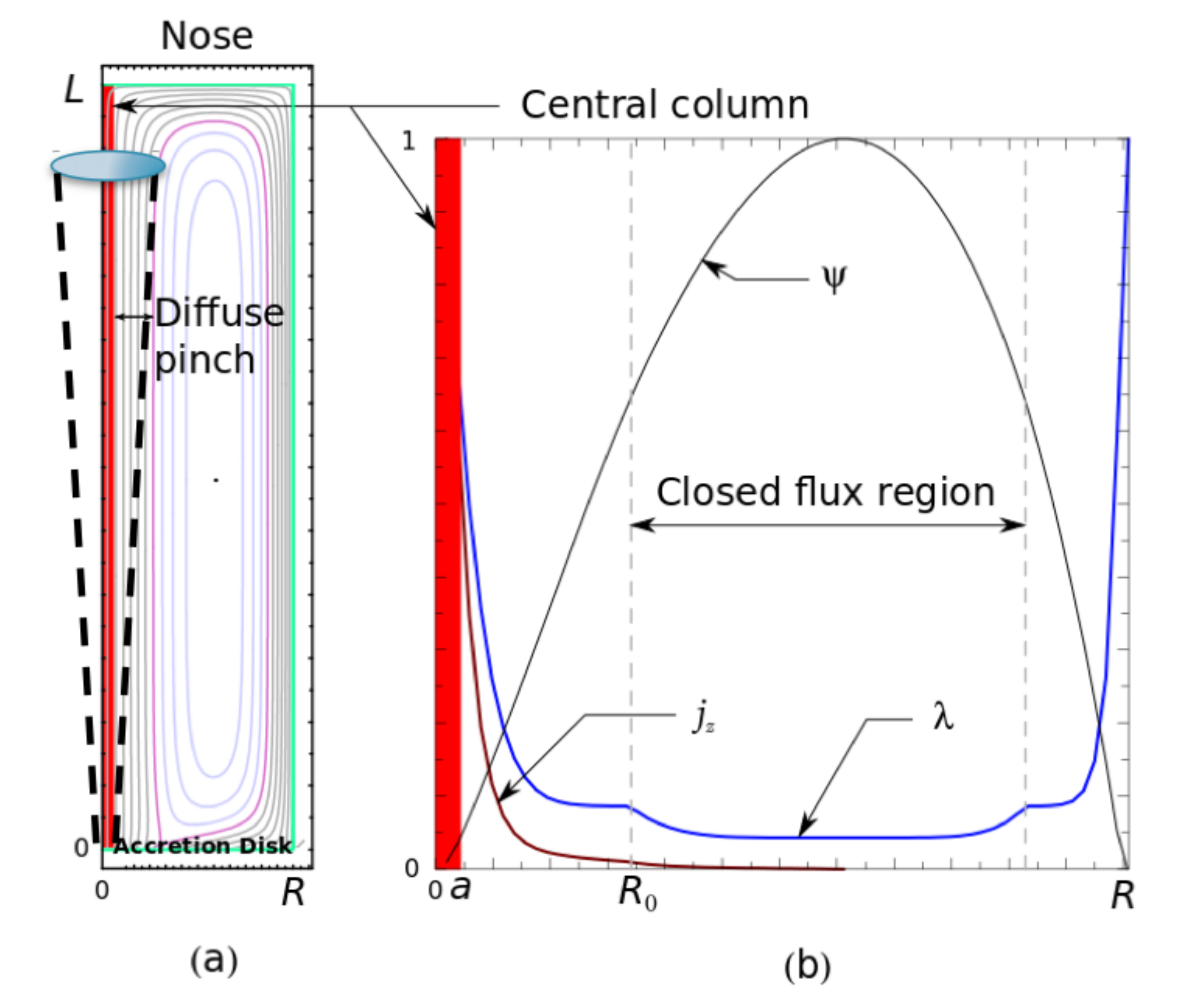} & 
\end{align}
\caption{Left: A simplified sketch of an accretion disk ejecting a magnetic jet of length $L$ and radius $R$, overlaid by a Grad-Shafranov solution for poloidal flux $\psi(r,z)$ (field $B_z,B_r$). Right: G-S boundary conditions derived in Papers I and II (with $\lambda(\psi)=|j_z/B_z|$). Note the concentration of current in a Central
Column (red), embedded inside a Diffuse Pinch in which outgoing flux
surfaces are straight, finally turning at $z = L$. The dashed cone concerns synchrotron radiation, 
Section \ref{sec:ele}.}\label{Fig1}
\end{figure}

This final structure is shown in Figure \ref{Fig1}, taken from Paper II, showing an $r$-$z$ cross-section of vertically-expanding magnetic flux surfaces that is one way to describe the jet produced by the spinning accretion disk at the bottom of the figure, and often the description provided by MHD simulations of disk/jet systems.
For our purposes it is more useful to note that the structure in Figure \ref{Fig1} 
is mainly produced by electric current flowing vertically up the Central Column, then bending to form the jet “nose” where the majority of
cosmic ray acceleration will take place. Following \citet{lynden-bell2003}, we will interchangeably refer to jets as the Central Column or as a ``magnetic tower." We postpone to Section \ref{sec:jetpara} to discuss the role of the weaker but important current surrounding the Central Column, labeled Diffuse Pinch in Figure \ref{Fig1}. 

\begin{figure}\nonumber
\begin{align}
  \hspace{0cm} \includegraphics[height=150pt,trim = 6mm 1mm 0mm 1mm]{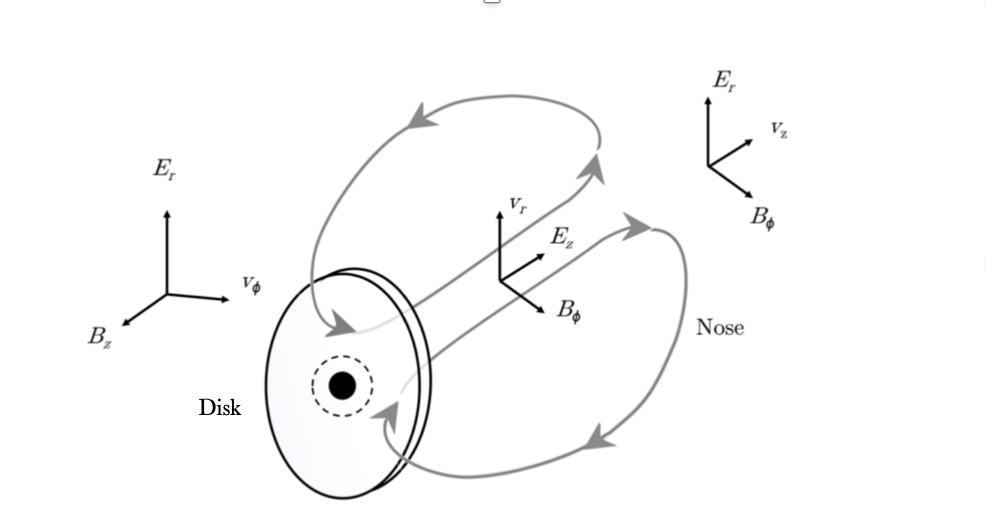} & 
\end{align}
\caption{Sketch of an accretion disk accelerator producing an axi-symmetric jet current $j_z$ of slowly growing length $L$, showing the poloidal component of two of the closed current loops driven by an electrostatic field $E_r$ due to disk rotation. The $E_r$ in the disk extracts energy from accretion. The
$E_r$ in the nose contains the cosmic ray accelerator.}\label{Fig2}
\end{figure}

As is shown in Figure \ref{Fig2}, the Central Column current finally loops through the disk, 
serving as an electric circuit extracting power from the disk. In our model, the Central Column current is ultimately self-collimated by the magnetic pinch force, 
$\mathbf{j}\times \mathbf{B}$, creating a magnetic tower (or ``pinch"). As noted in the Abstract, we have shown in \citet{fowler2016} -- hereafter Paper III -- that the existence of a self-collimated jet current automatically produces the electric fields that accelerate cosmic rays, as is further elaborated here in Section \ref{sec:ionacc}. 

But besides $\mathbf{j}\times \mathbf{B}$, a spinning accretion disk produces an electric field $\mathbf{E}$ that actually turns out to be the force that launches jets from an accretion disk \citep{mckinney2012}. Thus how jets launched by $\mathbf{E}$ evolve to magnetically-collimated ``towers" 
becomes a vital part of the story.

\subsection{How Magnetically-Collimated Jets Are Launched by an Accretion Disk}

Magnetically-collimated jets produced in the laboratory employ plasma guns driven by a capacitor bank \citep{hooper2012,zhai2014,fowler2016}. 
Accretion disk jets are driven by disk rotation $\Omega$,
creating a dynamo similar to a Faraday disk but with self-excited growing magnetic flux \citep{balbus1998}, 
given any pre-existing seed field near an evolving black hole (Appendix A.3).
We find that jets created by rotation can be fundamentally different from laboratory jets, yet they evolve to the same thing, but in a manner not easily accessible to GRMHD computer simulations of accretion disks. 

\begin{figure}\nonumber
\begin{align}
  \hspace{0cm} \includegraphics[height=200pt,trim = 6mm 1mm 0mm 1mm]{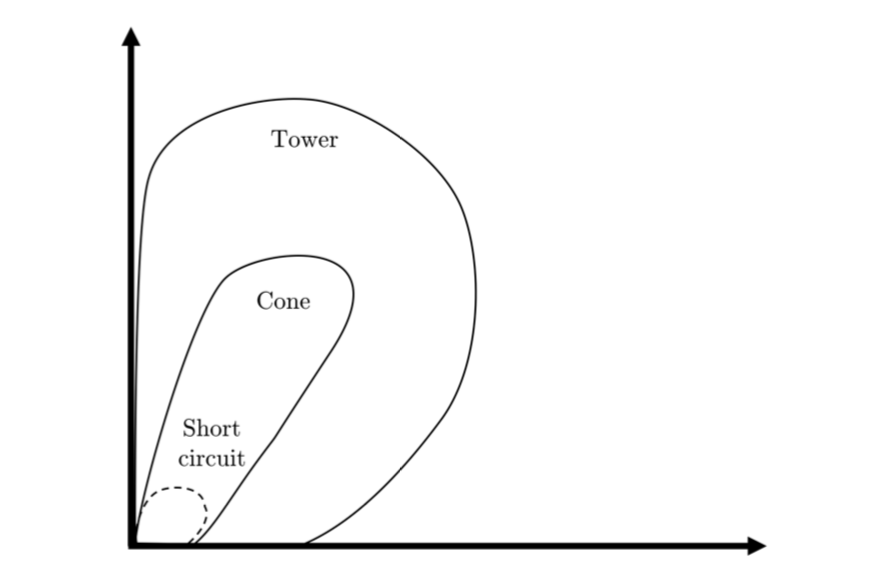} & 
\end{align}
\caption{Showing a conical jet embedded inside a magnetic tower. The magnetic tower return current acts as a barrier forcing the conical jet to slow down. Relativistic Alfv\'en waves reflecting from the tower return current serve mainly to smooth out the jet structure inside this moving barrier (see Paper II). 
The cone launched at speed $c$ can continue at speed $c$ temporarily due to the short-circuit current that maintains a constant jet current while the disk current continues to grow (see Figure \ref{Fig4}).}\label{Fig3}
\end{figure}

The main features of jet evolution occur in three steps:
\begin{enumerate}[(1)]
\item First the dynamo electric field $E_r = \left|\frac{r \Omega B_z}{c}\right|$ launches a jet at velocity $c$ when the dynamo voltage $V = \int dr E_r$ drives the dynamo current up to a value given by $I = \frac{V}{Z_o}$ with free-space impedance $Z_o = \frac{1}{c}$ in our units \citep{blandford2017}. 
This jet is not collimated but instead has the conical shape sketched in Figure \ref{Fig3} \citep{tchekhovskoy2008}.
\item Next, as the dynamo current continues to rise, the jet current remains fixed as long as the jet velocity persists at $\frac{dL}{dt} = c$. 
The growth in dynamo current greater than the jet current forces a portion of the dynamo current 
to short-circuit through the disk corona as shown at the bottom of Figure \ref{Fig3}. The short-circuit is produced 
by acceleration of the jet to speed $c$. 
\item Finally, anything that slows down $\frac{dL}{dt} \ll c$ eliminates the short-circuit, allowing the 
conical jet to evolve to a collimated magnetic tower with current equal to the full current produced by the dynamo. 
Why this last stage of jet evolution is not observed in GRMHD simulations is discussed in Section \ref{sec:evidence}. 
\end{enumerate}

Figure \ref{Fig3} again calls attention to the utility of describing jets as current rather than flux. The current $I$ in 
Step (1) above is the total current produced by the accretion disk acting as a self-excited dynamo creating its own magnetic field \citep{balbus1998}. The current $I$ is the maximum current in the system, referred to interchangeably as the disk current or the dynamo current. 

Loops in Figure \ref{Fig3} represent the main current paths emerging near the axis of symmetry and returning to the disk. Figure \ref{Fig3} shows a branching of the dynamo current $I$ as it emerges to form a jet. The initial branch forms the jet labeled Cone; followed by the short-circuit as the dynamo current $I$ grows while the jet current stays constant; 
and finally the Tower branch grows as the Cone dies so that the tower jet becomes the total current $I$ produced by the dynamo. 


\subsection{Dynamics of Accretion Disk/Jet Structures: Mathematical Model}

The conclusions in Section 2.2 can be derived from the following relativistic current and momentum equations, with gravitational potential $V_G$ to represent the black hole. General relativity included in GRMHD codes is not required except very near the black hole, leaving then special-relativistic fluid equations in three dimensions (3D) derived from Vlasov distribution functions $f(\mathbf{x},\mathbf{p},t)$ with position $\mathbf{x}$, momentum $\mathbf{p}$ and time $t$ in our reference frame. 
We obtain:
\begin{align}
\left[ \frac{\p \mathbf{j}}{\p t} + \text{K.E.} \right] &= \sum \int d\mathbf{p} \,\frac{e^2}{m\gamma_L} f \left[ \mathbf{E} - \frac{1}{c^2}\mathbf{u}\left( \mathbf{u} \cdot \mathbf{E} \right) + \frac{1}{c} \mathbf{u} \times \mathbf{B} \right]~, \tag{1a}\\
\frac{d\mathbf{P}}{dt} &= \left[ \frac{1}{c}\mathbf{j} \times \mathbf{B} + \sigma \mathbf{E} - \rho \nabla V_G - \nabla p_{\text{amb}} \right]~. \tag{1b}
\end{align}
Here $\mathbf{j} = \sum \int d\mathbf{p} \, q \mathbf{u} f$ is the current density where $\sum$ sums over ions (hydrogen) and electrons with charge $q = \pm e$, rest mass $m$; $\mathbf{P} = \sum \int d\mathbf{p} \, \mathbf{p} f$ is the bulk flow
momentum; $\sigma$ is the charge density; $\text{K.E.}$ represents small kinetic terms; and $\mathbf{u} = \left(\mathbf{p}/m \gamma_L\right)$ is the particle velocity giving for example a relativistic correction $\vec{u}(\vec{u}\cdot\vec{E})$ arising from $\frac{\partial \gamma_\mathrm{L}}{\partial p}$ in an integration by parts 
\citep{montgomery1964}. 

Gravity $V_G$ in Equation (1b) divides the system into a denser accretion disk and lower density jet ejected in the disk corona. 
We obtain mean-field (2D) equations for the jet by averaging over $\phi$ and fluctuation scales in time and space. 
Ambient pressure   $p_{\text{amb}}$ represents ionized gas outside the accretion environment.
Dropping the ambient pressure $p_{\text{amb}}$ for the moment, 
we obtain:
\begin{align}
\mathbf{E} + \frac{1}{c} \mathbf{v}_{\perp} \times \mathbf{B} &= - \frac{1}{c} \frac{\p \mathbf{A}}{\p t} - \nabla \Phi + \frac{1}{c} \mathbf{v}_{\perp} \times \mathbf{B} = \mathbf{D} \tag{2a} \\
\frac{d\mathbf{P}}{dt} &= \frac{\p \mathbf{P}}{\p t} + \nabla \cdot \sum \int \, d\mathbf{p} f \left(\frac{\mathbf{p}\mathbf{p}}{m \gamma_L}\right) = \frac{1}{c}\mathbf{j} \times \mathbf{B} + \sigma \mathbf{E} = \frac{1}{c}\mathbf{j}_{\perp}^* \times \mathbf{B} \tag{2b} \\
\sigma &= \frac{\nabla \cdot \mathbf{E}}{4\pi} \tag{2c} \\
\mathbf{j}_{\perp}^* &\equiv \mathbf{j}_{\perp} - \sigma c \left(\mathbf{E} \times \frac{\mathbf{B}}{B^2}\right) = -\frac{c}{B^2}\left(\frac{d\mathbf{P}}{dt} \times \mathbf{B} \right) \tag{2d} \\
\left(\frac{1}{c}\mathbf{j} \times \mathbf{B} + \sigma \mathbf{E}\right)_r &= \frac{1}{c} j_{\phi} B_z - \frac{1}{8\pi r^2}\frac{\p}{\p r} r^2 \left({B_{\phi}}^2 - {E_{r}}^2\right) = 0 \tag{2e} 
\end{align}
Equation (2a) is Ohm's Law obtained by dropping terms in Equation (1a) as discussed in Section 3, with fluid velocity $\mathbf{v}$, vector potential $\mathbf{A}$, electrostatic potential $\Phi$ and hyper-resistivity $\mathbf{D}$ representing any kind of turbulence, MHD or kinetic \citep{fowler2007}. 
Equation (2b) is the momentum equation, with $\gamma_L = \sqrt{1 + \left(\frac{p}{mc}\right)^2}$, and 
it includes the electric force due to space charge density $\sigma$, given in Equation (2c). 
Equation (2d) is the solution to Equation (2b) giving $\mathbf{j}_{\perp}^*$ as acceleration-driven current flow between flux surfaces moving at velocity $\mathbf{v}_{\perp} = c \left(\mathbf{E} \times \mathbf{B}/B^2\right)$ in our reference frame. 
Equation (2e) is the Force Free Degenerate Electrodynamics (FFDE) approximation to Equation (2b) \citep{meier2012},
assuming that hyper-resistivity $\mathbf{D} = 0$ . 

\subsection{Step 1: The Formation of Conical Jets}

We now apply Equations (2a) - (2e)   to see how theory corresponds to the three steps of jet evolution 
described in Section 2.2, from Cone to Short Circuit to Magnetic Tower. Three distinct regions of the disk/jet system will emerge associated with three different timescales within the overall lifetime of the jet ($\tau \approx 10^8$ yrs):
\begin{enumerate}[(1)]
\item The disk itself: where the dynamo current reaches quasi-steady state in a few accretion times (only years near the black hole);
\item  The disk corona: home of a persisting short-circuit that allows continuation of a conical jet for a duration of
$0.01\tau= 10^6$ yrs (see Section \ref{sec:evidence});
\item The jet as magnetic tower: enduring for the remaining $0.99 \tau$.
\end{enumerate}
It is the great disparity in timescales that makes it difficult to span the entire evolution of a magnetic tower 
in GRMHD disk+jet simulations which focused on the early phases of dynamo formation; or in special 
relativistic MHD simulations which 
focused on jet propagation.   

\subsubsection{Initial Conditions} 
\label{sec:initi}

As in many GRMHD simulations, we consider an initial state in which a pre-existing black hole is embedded 
in an accretion disk with rotation frequency $\Omega$, threaded by a vacuum poloidal seed field $B_z$ 
created by distant currents in the ambient. Both dipole and quadrupole seed fields eject jets upward 
and downward, the upward jets being similar except very near the disk midplane 
(see Paper I, Appendix B). Figures \ref{Fig1} - \ref{Fig3} depict an upward jet. 

In the absence of MRI, the ideal Ohm's Law only allows 
gravity to compress flux, sometimes observed in simulations. Here we are interested in the formation 
of a self-excited dynamo that grows its own magnetic field to the kilogauss levels inferred from 
AGN observations (Paper I, Balbus \& Hawley 1998).  
Rotation immediately causes an accretion disk to charge up to 
$E_r = - (r\Omega/c)B_z$ (positive for our sign convention) with a charge density
$\sigma = (1/4\pi r)\p(rE_r/\p r)$. Because the ordinary resistive conductivity is negligible in disks, 
no dynamo current flows through the disk until the MRI instability sets in, creating hyper-resistive current flow 
radially across flux surfaces inside the disk (as ordinary resistivity allows inside a Faraday disk). 
Closure of this MRI-driven current in or above the corona allows a jet to be ejected.

\subsubsection{Two Quasi-Steady States} 

We note that, for the quasi-steady state described by Equation (2e), there are two jet solutions, one with
$E_r = B_\phi$ and $j_\phi = 0$; and another with $E_r = 0$ giving ${\bf j}\times {\bf B} = 0$.  
Since initially both $j_\phi$ and $j_z$ are zero in the vacuum seed field, before a jet is launched 
${\bf j}$ is zero, giving, by Equation (2b), jet acceleration $\frac{d\mathbf{P}}{dt} = \sigma \mathbf{E}$ 
that saturates at $\frac{1}{c}\, \mathbf{j} \times \mathbf{B} = -\sigma \mathbf{E}$. 
In Equation (2e), this corresponds to  
$j_z = c \sigma \left( \frac{E_r}{B_{\phi}}\right)$, possible if $c \sigma =  \frac{c}{4 \pi r} \frac{\p}{\p r} \left( r E_r \right) = \frac{c}{4 \pi r} \frac{\p}{\p r} \left( r B_{\phi} \right) = j_z$ with solution $E_r = B_{\phi}$ as the threshold to launch a jet. 
This is the conical jet solution in GRMHD simulations \citep{mckinney2012},
conical rather than collimated because $E_r$ cancels the $B_\phi$ pinch force. 
A second solution exists, however, when $E_r \ll B_\phi$. 
In this limit, a jet can be launched by $d\mathbf{P}/dt  = c^{-1} {\bf j}\times {\bf B}$ 
which gives jets as magnetic towers in the laboratory (Hooper et al. 2012; Paper II). Finally, 
we call attention to the kinetic term in $d\mathbf{P}/dt$  that is dropped in the FFDE 
approximation but plays a vital role in sustaining a short circuit in the disk corona, as discussed in Section 
\ref{sec:shortC}.

\subsubsection{Which Comes First -  Cone or Tower?} 

As the dynamo current builds up, the conical jet comes first because more dynamo current is required to launch a tower jet than a cone. This is shown in Figure \ref{Fig4}, which plots the cumulative jet current
$I(r) = \int_0^r 2 \pi r \, dr j_z$ in units of $I_A$, where $I_A = \int_0^a 2 \pi r \, dr j_z$ with $a = 10 R_s$  
(Schwarzschild radius). 
The shape of $I(r)$ is only approximate at $r < a$. But for $r > a$,  
the curves are calculated from the radial force equation with angular momentum as a constraint, 
given by $(1/2) \dot{M} \Omega(r) = r|B_z| B_\phi$ with Keplarian rotation, in Equation (6c) below. 
The lower curve for the conical jet gives $I = (rB_\phi c/2)$ with $B_\phi = E_r = (r\Omega/c) |B_z|$ from force
balance in Equation (2e) with $j_\phi = 0$, giving $B_\phi \propto \Omega(r) \propto (a/r)^{3/2}$. 
The upper curve for the tower is taken from Figure 2 in Paper I, obtained by an exact solution of 
$({\bf j}\times {\bf B})_r = 0$ with the angular momentum constraint. The total current asymptotes to a 
constant value. Note that $I(r)$ for the cone peaks at $r = a$, indicating a reversal in sign of the jet current density 
noted in simulations, while, for the tower, $r = a$ is merely the radius of the Central Column. 
The value of $a$ and the ratio of cone and tower currents at $r = a$ are derived in Section \ref{sec:jetpara}.

\subsubsection{Continued Growth of the Disk Current} 

Next we consider the timescale of processes inside the disk. The growth of the disk current in terms of the vector potential is described by Equation (2a) with ${\bf D}$ equal to that for MRI. Near the black hole, the growth rate 
$(cD/A)$ 
will turn out to be of the order of the accretion rate $(v_r/a)$, giving a rapid 
buildup of dynamo current to balance gravitational energy input by accretion (only years, accessible by GRMHD, as noted above). We obtain inside the disk: 
\begin{align}
\frac{\p A_r}{\p t} - \left(v_{\phi} B_{z} - c \frac{\p \Phi}{\p r}\right)  & \approx \frac{\p A_r}{\p t} \approx -\left( v_z B_{\phi} + c D_{r} \right) \tag{3a} \\
\frac{\p A_{\phi}}{\p t} &= -\left(v_r B_z - v_z B_r + c D_{\phi}\right) \tag{3b}\\
c \frac{\p \Phi}{\p r} &= v_{\phi} B_z \tag{3c} \\
\frac{E_r}{B_{\phi}} &\approx \left(-\frac{\p \Phi}{\p r}\right)/B_{\phi} = -\frac{v_{\phi}}{c}\left(\frac{B_z}{B_{\phi}}\right) = \frac{v_{\phi}}{c}\left(-\frac{D_{\phi}}{D_r}\right)\left(\frac{v_z}{v_r}\right) \ll 1 \tag{3d} \\
-\frac{D_{\phi}}{D_r} &\approx 1 \tag{3e}~~.
\end{align}

\noindent We note that, as the disk develops MRI, these equations show that $B_\phi \approx {\p A_r}/{\p z}$ 
gradually grows from zero, until reaching $B_\phi = E_r$ which launches the conical jet, but then continuing smoothly 
to the value needed to sustain a magnetic tower. 
It is this gradual growth of $B_\phi$ inside the disk, independent of the jet, that led us to postulate the 
creation of a short-circuit in the corona isolating the growing current in the disk from a constant 
current in the jet. Though initially $B_\phi < E_r$,  it grows smoothly to equal and surpass $E_r$,
as discussed below.

\begin{figure}[H]\nonumber
\begin{align}
  \hspace{0cm} \includegraphics[height=200pt,trim = 6mm 1mm 0mm 1mm]{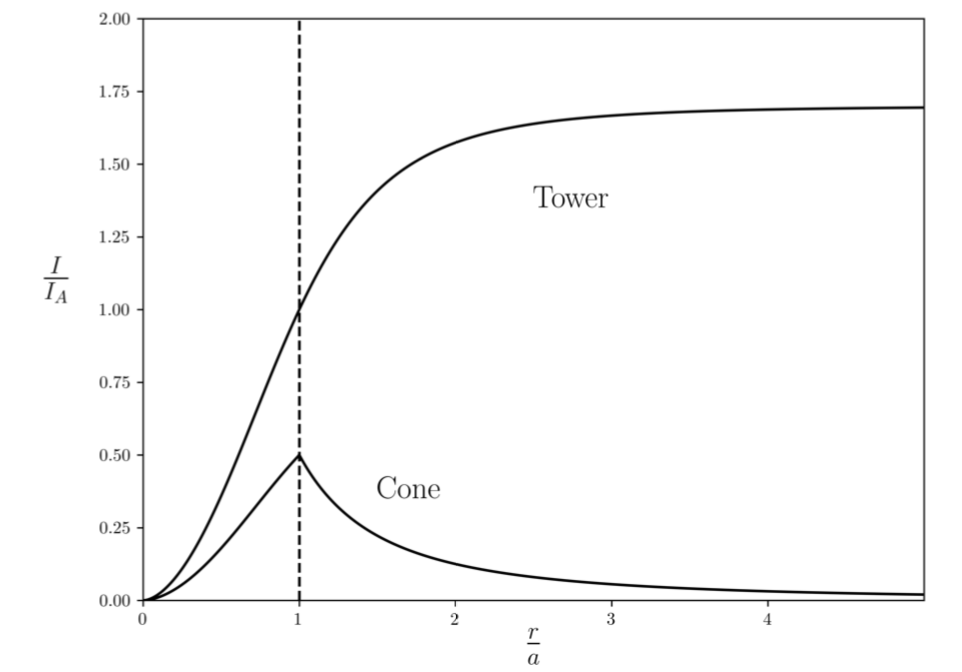} & 
\end{align}
\caption{Threshold current $I(r) = rB_\phi c/2$ profile for onset of a conical jet (bottom curve) and a magnetic tower jet (upper curve, from Paper I).}\label{Fig4}
\end{figure}

In Equations (3a) and (3b), we apply Equation (3c) to eliminate zeroth-order terms, 
leaving $\frac{\p {\bf A}}{\p t} \approx - c\mathbf{D}_{\text{MRI}}$ as the driver of the self-excited 
dynamo magnetic field. Then 
Equation (3d) gives the steady state  with $E_r \ll B_\phi$ obtained as follows. 
 We set ${\p A_r}/{\p t} = {\p A_\phi}/{\p t} = 0$ and drop $v_z B_r$ compared to 
$v_r B_z$ inside the disk, giving then $B_\phi = -(cD_r/v_z)$ and $B_z \approx - (c D_\phi/v_r)$ in 
quasi-steady state. And most importantly, we apply Equation (3e), 
derived from an ordering scheme in Appendix A of Paper I, and reaffirmed for an MRI-driven $\mathbf{D}$ in Appendix A of this paper (giving $\frac{|B_z|}{B_{\phi}} = \frac{v_z}{v_r} = \left(\frac{a}{r}\right)^{1/2}$ in  Paper I). 
Equation (3d) shows why $B_{\phi}$ in the disk grows to exceed $E_r$ in the disk, enough so to drive the jet current $I(r)$ up to the magnetic tower threshold in Figure \ref{Fig4}.

\subsection{Step 2: The Short Circuit}
\label{sec:shortC}

Why a magnetic tower does not form immediately concerns the persistence of the initial conical jet. 
This is due to the jet structure whereby the jet current returns close to the outgoing jet so that jet magnetic energy only changes as the length $L$ changes, at a rate $\frac{dL}{dt} = c$ (Paper II, Appendix C; \citet{blandford2017}). 
As was noted in Step (2), Section 2.2, a short-circuit is required to sustain a constant jet current as the dynamo current rises. The required short-circuit is automatically accommodated by $\mathbf{j}_\perp^*$ in
Equation (2d). This occurs through the kinetic term of $d\mathbf{P}/dt $ that allows a quasi-steady $(\partial\mathbf{P}/\partial t = \mathbf{0})$ conical jet to persist even as the disk current continues to grow, giving then the short circuit current $I_\perp$ and a corresponding parallel current $I_{||}$ as follows: 
\begin{align}
I_{\perp} &= 2 \pi r \Delta_z j_{\perp r}^* = \left(\frac{c r B_{\phi}}{2}\right)_{r=a}  \sum \int d\mathbf{p} \, f \left(\frac{4 \pi {p_{\perp z}}^2}{m \gamma_L B_\phi^2}\right) \approx \frac{1}{2}I\left(\frac{v_{\perp z}}{v^*_A}\right)^2 \tag{4a} \\ 
I_{\parallel} &= I - 2 I_{\perp} = I \left( 1 - \frac{{v_{\perp z}}^2}{{v^*_A}^2}\right) \tag{4b} \\
E_r &= \frac{1}{c}\left(r \Omega |B_z| + v_z B_{\phi}\right) \tag{4c} \\
v^*_A  &\approx  c \left[\frac{B_\phi}{\sqrt{B_\phi^2 + 4 \pi n m c^2}}\right] \tag{4d}\\
I ({\rm disk}) & = I_\parallel  + I_\perp ({\rm corona}) + I_\perp ({\rm nose}) \tag{4e}
\end{align}
Here $\Delta_z$ is the width of the current path and the factor $\left(\frac{c r B_{\phi}}{2}\right)_{r=a} = I$ 
is the final disk current with factor $\frac{1}{2}$ fitted to Figure \ref{Fig4} at $r=a$.
Equation (4a) is obtained from Equation (2d) with $\frac{d\mathbf{P}}{dt}$ from Equation (2b) with $\frac{\p\mathbf{P}}{\p t}= \mathbf{0}$. 
The cross-product of $\left(\frac{d\mathbf{P}}{dt} - \frac{\p \mathbf{P}}{\p t}\right)$ with $\mathbf{B}$ 
gives $p_{\perp z} B_{\phi}$; and maximizing on the divergence gives 
$\Delta_z \left(\nabla \cdot \mathbf{p}\right) = p_{\perp z} = m\gamma_L u_{\perp z}$. 
The final expression approximates the integral over $u_{\perp z}$ by the fluid velocity $v_{\perp z}$
divided by the Alfv\'en-like velocity $v^*_A$ in Equation (4d). 
Equation (4c) is Ohm's Law 
in the conical jet, whereby jet acceleration to $v_{\perp z} \approx c$ maintains $E_r = B_{\phi}$ in the jet.

\subsubsection{Parallel Current, Circuit Notation}

The distinction between $I_{\perp}$ and $I_{||}$ in Equations (4a) and (4b) reflects the branching 
of the total jet current $I$ in Figure \ref{Fig3}. 
While the MRI-driven current loop for a magnetic tower must close along closed poloidal field lines, 
the current loop of the initial conical jet closes via a current $I_\perp$ 
flowing directly across poloidal flux surfaces. This $I_\perp$  
flows across the nose of the growing jet. But because the conical jet current remains constant as 
the disk current continues to increase, 
a short-circuiting $I_\perp$  must also flow in the corona, giving the disk current as the sum of 
three branches in Equation (4e). We take the conical jet current equal to its return current at the 
nose so that Equation (4e) gives $I_\parallel = I - 2 I_\perp \rightarrow I$ in Equation (4b).

In the next Section, we explicitly describe Figure \ref{Fig3} as an electric circuit. 
Here and hereafter, the circuit variables $I$ and $V$ are defined as follows. 
The symbol $I$ will be the disk or dynamo current, the largest current in the system, 
ultimately equal to the jet current in a self-collimated magnetic tower. 
As in Figure \ref{Fig4}, $I = I_A$, the current inside the Central Column 
$(r \leq a)$.  The symbol $V$ will denote the dynamo voltage, defined as the voltage 
across the Central Column where it intercepts the disk, as in Step (1), Section 2.2, 
giving $V=\int_{R_S}^a dr E_r$.

\subsection{Step 3: Slowing Down of the Jet, Cones to Towers}

As in Paper II, we describe jet propagation by an electric circuit satisfying energy conservation. We obtain: 
\begin{equation}
\frac{\p}{\p t}\left[ \frac{B^2 + E^2}{8 \pi} \right] + \mathbf{v} \cdot \frac{d\mathbf{P}}{dt} + \nabla \cdot \left( \frac{c \mathbf{E} \times \mathbf{B}}{4\pi} \right) = - \mathbf{v} \cdot \rho \nabla V_G \tag{5a}
\end{equation}
\begin{align}
f_{\text{dis}}(IV) = \frac{d}{dt} \int d\mathbf{x} \, \left( \frac{B^2 + E^2}{8 \pi}\right) &\approx \frac{d}{dt}\left\{L \left(\frac{I}{c}\right)^2 \left[ 1 + \frac{{I_{\perp}}^2}{I^2}\ln\left(\frac{R_2(z)}{R_1(z)}\right) + \frac{{I_{\parallel}}^2}{I^2} \ln\left(\frac{R}{a}\right)\right]\right\} \tag{5b} \\
\frac{dL}{dt} &\approx \frac{f_{\text{dis}}\,c}{1 + \frac{1}{4}\frac{{v_{\perp z}}^4}{{v^*_A}^4} \ln\left(\frac{R_2(z)}{R_1(z)}\right) + \left(1 - \frac{{v_{\perp z}}^2}{{v^*_A}^2}\right)^2 \ln\left(\frac{R}{a}\right)} \tag{5c} \\
\frac{dL}{dt} &\rightarrow \frac{f_{\text{dis}}\,c}{1 + \ln\left(\frac{R}{a}\right)} \quad ; \quad v_{\perp z} \ll v^*_A \tag{5d} \\
f_{\text{dis}} &= \frac{1}{2}(1 - f_{\text{conv}}) \tag{5e}~~.
\end{align}
Equation (5a) is obtained in the usual way by adding the results from dotting $\mathbf{v}$ into Equation (2b) and dotting $\mathbf{B}$ into Maxwell's $\frac{\p \mathbf{B}}{\p t}$ equation, using Equation (2a) giving $\mathbf{v} \cdot \mathbf{E} = 0$ if we neglect $\mathbf{D}$ in the jet. 
Integrating Equation (5a) over the jet volume gives Equation (5b), with dynamo power on the left hand side (from the Poynting vector in Equation (5a)), and in Equation (5e) a dissipation factor $f_{\text{dis}}$ with $f_{\text{conv}}$ as the efficiency of converting magnetic energy to ion acceleration and a factor $1/2$ representing ambient shocks, as in Paper II. 
As in Figure \ref{Fig3}, Equation (5b) approximates the jet as a conical jet with current $I_{\perp}$ embedded inside a magnetic tower with current $I_{\parallel}$. 
We drop $E^2$ and divide $\int d{\bf x} B^2$ into three parts: (a) the Central Column giving $1$ 
in the bracket $[…]$; (b) the conical jet giving the term $\propto  I_\perp^2$; and 
(c) the enveloping magnetic tower gives the term $\propto I_\parallel^2$, valid when $j_\parallel >  j_\perp$, 
causing current to twist around field lines. As in Paper II, we use the fact that, as $dL/dt$ falls below $c$, 
Alfv\'en waves at speed $c$ (for the jet density in Section \ref{sec:jetpara}) spread flux radially to 
produce a blunt nose like Figure \ref{Fig1}, whereby the jet can be approximated as a cylinder of 
fixed radius and expanding length $L$ approximately independent of $r$ so that $L$ can be removed 
from the integral to give Equation (5c).

Equation (5c) derives the jet velocity $\frac{dL}{dt}$ from Equation (5b) in quasi-steady state (constant
$I$), using Equation (4a) for $I_{\perp}$ and Equation (4b) for $I_{\parallel}$.  
The slow growth of the jet current to match the final dynamo current is approximated by $cV/I = 1$ giving the correct limits both as the jet slows down and when the jet is first launched with jet current $I = V/Z_o = cV$. 

Formulating jet propagation as an electric circuit has the advantages that the black hole region omitted in Equation (1b) is included in the circuit 
\citep{mcdonald1982,frank2002},
and the magnetic tower can be solved analytically at $r > a = 
10 R_S$, giving $B_{\phi} \propto 1/r$ yielding the logarithmic factor $\ln\left(R/a\right)$ in Equation (5b) (Papers I \& II). 
That $B_{\phi} \propto 1/r$ also holds inside the conical jet, giving $\ln\left(R_2/R_1\right)$ for a cone with outer and inner radii $R_2$ and $R_1$, can be derived from the $r$-component of Ohm's Law in Equation (2a), with $D_r \approx 0$ in the jet. We obtain:  
\begin{align}
{B_{\phi}}^2 = {E_r}^2 &= {B_{\phi}}^2 \left(\frac{v_{\perp z}}{c}\right)^2 + 2 \left(\frac{r \Omega v_{\perp z}}{c^2}\right)\left(|B_z|B_{\phi}\right) + \left(\frac{r \Omega}{c}\right)^2 B_z^2 \tag{6a} \\
B_{\phi} &= \Omega \left(\frac{\dot{M}/2c}{1 - v_{\perp z}/c}\right)^{1/2} \approx \frac{\sqrt{\dot{M}c}}{r} \tag{6b} \\
\frac{1}{2}\dot{M}\Omega\left[1-\frac{3}{2}\frac{\nu}{rv_r}\right]\approx\frac{1}{2}\dot{M} \Omega &= \int_0^z dz \left(\frac{4 \pi r}{c}\right)\left(\mathbf{j} \times \mathbf{B}\right)_{\phi} =  \left[r |B_z|B_{\phi}\right]_{z=H} \qquad \text{ at disk} \tag{6c} \\
\frac{v_{\perp z}}{c} &=  \sqrt{ 1 - \left[\left(\frac{r \Omega}{c}\right)^2 + \left( \frac{1}{{\gamma^*_L}^2} \right)\right]} = 1 - \frac{1}{2} \left(\frac{r \Omega}{c}\right)^2 + ... \tag{6d}
\end{align}
The right hand side of Equation (6a) is the square of Equation (4c), with Keplerian rotation $\Omega = \left(M G / r^3\right)^{1/2}$ for black hole mass $M$ and Newtonian gravitational constant $G$. 
Solving Equation (6a) for $B_{\phi}$ gives exactly the  middle expression in Equation (6b) using angular momentum conservation discussed in Paper II, giving Equation (6c) with fields defined at the disk corona ($z = H$) and accretion rate $\dot{M} = \frac{dM}{dt}$. 
We approximate the relativistic Lorentz factor  by 
$\gamma^*_L = [1 - \frac{1}{c^2}({v_{\perp z}}^2 + r^2 \Omega^2)]^{-1/2}$ with fluid velocities $v_{\perp z}$ 
and $r\Omega$. Then solving for $v_{\perp z}$ and expanding gives Equation (6d), valid if 
$1/\gamma^{*2}_L  < (r\Omega/c)^2$, true at all $r > a$ with $\gamma^*_L > 5$ for numbers below.
Substituting Equation (6d) into Equation (6b) gives $B_{\phi} \propto \frac{1}{r}$ for the conical jet. 

That viscosity $\nu$ can be neglected in Equation (6c) is key to our model, whereby disk/jet angular momentum conservation governs all model predictions of jet properties and UHE cosmic rays and synchrotron radiation in Table 1 below. In Appendix A of Paper I, we derive viscosity as $\nu/(r v_r)=(k_zv_A/\Omega)(v_{1\phi}/v_{1r})\approx|v_r/v_\phi|^{1/2}$ giving 
$\nu/(r v_r)  \ll 1$ for our consistent ordering  and equipartition of turbulence $(v_{1\phi}\approx v_{1r})$ confirmed for MRI in Equation (A2.b) as noted earlier. (These conclusions are true at all $r>a$ despite an error in Paper I in estimating $v_{1\phi}$ as a projection perpendicular to $\mathbf{B}_0$ that underestimates $v_{1\phi}$ at large $r$.)

\subsubsection{How Cones Become Towers}

As anticipated in Appendix C of Paper II, it is the slowing down of the jet by Equation (5c) that yields the magnetic tower jet structure in Figure \ref{Fig1} giving the accelerator in Section 3. 

In Paper II, this was shown by following the dynamical history of $E_r/B_{\phi}$, given here by Equation (4c) yielding $\frac{dL}{dt} \approx v_{\perp z} \approx c(E_r/B_{\phi})$. 
Then anything that slows down the jet---dissipation and/or induction---implies $E_r \ll  B _{\phi}$ in Equation (2e). 
Extended to two dimensions, Equation (2e) with $E_r \ll  B _{\phi}$ becomes the Grad-Shafranov magnetic tower solution in Figure \ref{Fig1}, as derived in Paper II. 

\subsubsection{Eliminating the Short-Circuit}

The new result in this paper is the explicit origin of the short circuit current 
$\mathbf{j}^*$ by jet acceleration, in Equation (2d), as is required for the conical jet current to close on itself, in  
Figure \ref{Fig3},  while the parallel current producing a magnetic tower must follow poloidal flux surfaces. 

That slowing down dissipates $\mathbf{j}^*$ can be shown by re-writing Equation (2b), approximating ${\frac{d\mathbf{P}}{dt} \approx \rho \frac{\p \mathbf{v}}{\p t}}$ with $\mathbf{v}$ from Ohm's Law. 
We obtain: 
\begin{align}
\frac{d\mathbf{P}_{\text{POL}}}{dt} &\approx \rho \frac{\p}{\p t} \left(c \mathbf{E}_{\phi} \times \frac{\mathbf{B}}{B^2}\right) = \frac{1}{c} \left(\mathbf{j}_{\perp}^* \times \mathbf{B}\right)_{\text{POL}} - \nabla p_{\text{amb}} \tag{7a} \\
\frac{1}{c} \frac{\p E_{\phi}}{\p t} &\approx \left(\frac{dL}{dt}\right)^2\left(\frac{1}{c^2r}\right)\left(\frac{\p^2 \psi}{\p z^2}\right) \notag \\
&\approx	\left(\frac{4\pi j_{\phi}^*}{c}\right) - 4\pi r \frac{dp_{\text{amb}}}{d\psi} \tag{7b}
\end{align}
where we now retain the effect of ambient back pressure in $f_{\text{dis}}$ in Equation (5e). 
Equation (7b) is obtained by factoring out $\bm{\hat{\varphi}} \times \mathbf{B} = \frac{1}{r} \nabla \psi$ with 
poloidal flux $\psi = r A_{\phi}$ , with $\left(\rho c^2/4\pi B^2\right)=\left(c^2/v_A^2\right)=1$ 
for jet densities in Section \ref{sec:jetpara}.
Then slowing down $\frac{dL}{dt}$ eliminates $\mathbf{j}^*$ for typical ambient pressures over most of the nose surface, even though the integral effect of applying a weak back pressure over the stiff, blunt nose in Figure \ref{Fig1} does slow down $\frac{dL}{dt}$ by 50\% in Equation (5e), as discussed in Paper II.

The purely inductive rate of slowing down a conical jet when ambient pressure and dissipation are negligible is difficult to estimate. If the conical jet were a straight cylinder in a constant $B_z$ field, $B_z$ would never change; $j_\phi$ would remain zero; and the conical jet solution of Equation (2e) giving $B_\phi= E_r$ would persist. Given $B_\phi= E_r$, both the radial pinch force creating a magnetic tower and the hoop force causing growth  of the radiolobe radius $R$ are zero. The actual growth of inductance is associated with the 2D structure of the conical jet in Figure 3. We have not been able to calculate this 2D structure analytically, and there is reason to believe its radial profile cannot be in steady state. For example, the base of the jet should grow at the accretion rate, while GRMHD solutions $>10^4 R_S$ in length run for only a few accretion times.

\subsection{Evidence of Magnetic Collimation: Simulations and Observations
} 
\label{sec:evidence}

As already noted, attempts to simulate the entire astrophysical jet formation cycle are limited by the extreme range of time and space scales, from accretion disk dynamos concentrated near the black hole, where conical jets are created, to the Mpc dimensions of fully developed jets. In particular, GRMHD codes yielding MRI-driven dynamos producing conical jets in a low density ambient do not yet exhibit the slowing down required to produce a tower \citep{tchekhovskoy2017}. That this is perhaps understandable given the limited timescale of these simulations follows from Equation (5c) dropping ln(R/a) to obtain, for zero dissipation $(f_\mathrm{dis}=1),\ dL/dt\approx c/[1+\frac{1}{4}\ln (R_2/R_1)]>0.8c$ for typical cone dimensions in GRMHD 
simulations. On the other hand, assuming continued inductive slowing down of the jet for reasons discussed at the end of Section 2.6.2, as dissipation develops 
eventually giving $f_\mathrm{dis}=1/4$ in later sections, $I_{||}=(15/16)I$ yielding $dL/dt\approx cf_\mathrm{dis}/[1+0.9\ln(R/a)]$ as in Equation (5d). Equation (5d) predicts early slow-down to $dL/dt\approx 0.01c$ at 1$\%$ of the final jet length if we approximate $R(t) = 0.1L(t)$ in Equation (5d), indicating magnetic collimation over most of the life of the jet.

The slowing down by ln $R/a$ that produces a magnetically collimated tower is consistent with MHD simulations of magnetically-dominated jets, relativistic in \citet{guan2014}, non-relativistic in \citet{nakamura2006}; \citet{carey2009}; and \citet{carey2011}. Disk-like creation of jets by rotation of a conducting sphere in a dense ambient is featured in relativistic MHD simulations in \citet{bromberg2016}, exhibiting jet slowing down by the ambient and the two classes of MHD kink modes used in our accelerator model \citep{nakamura2006, nakamura2007, tchekhovskoy2016}. 

\begin{figure}[htb]
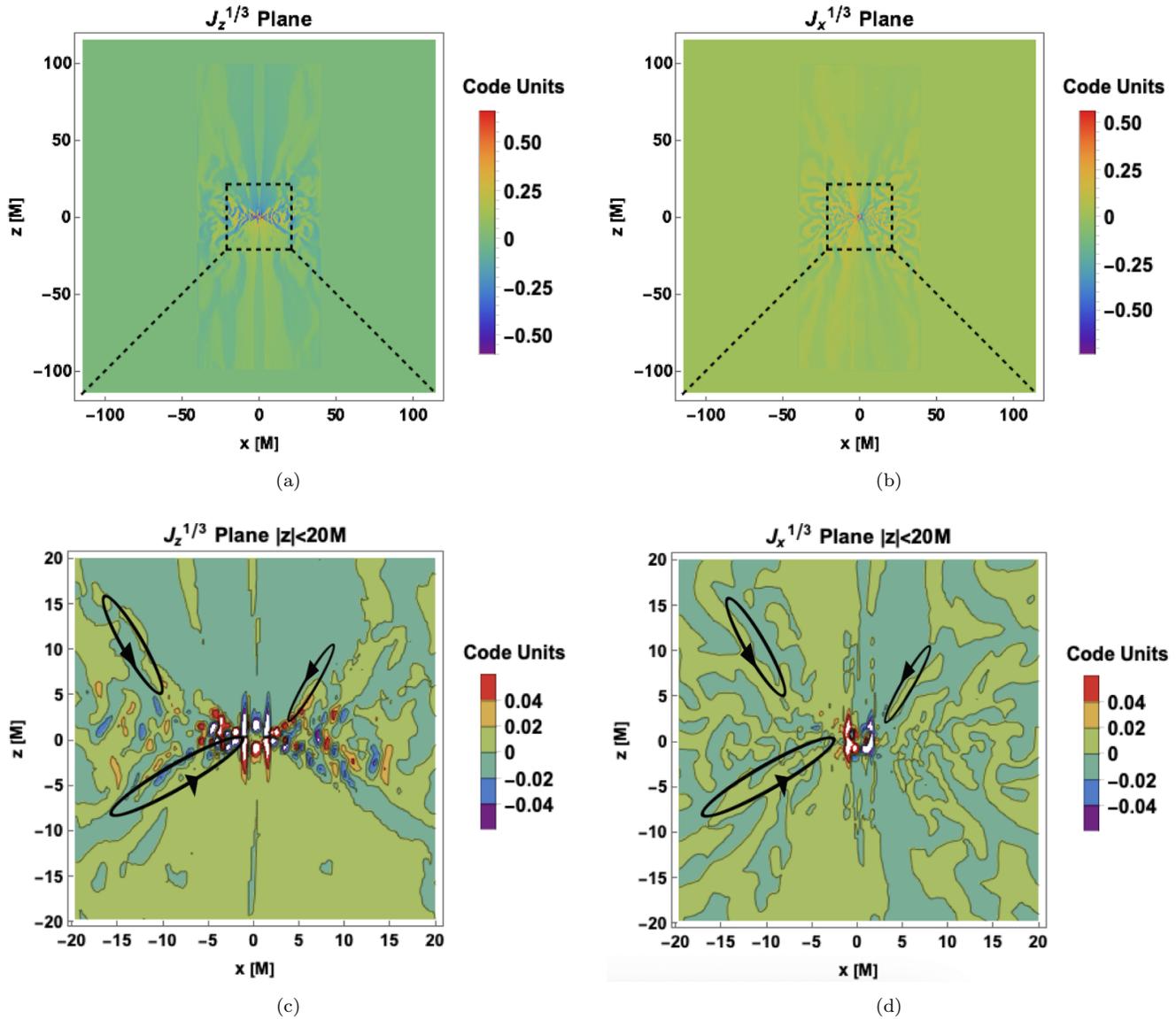

\gridline{\fig{FowlerLiAnantuaFig5a
.png}{0.47\textwidth}{(a)}  \fig{FowlerLiAnantuaFig5b
.png}{0.47\textwidth}{(b)}
          }
\gridline{\fig{FowlerLiAnantuaFig5c
.png}{0.47\textwidth}{(c)} \fig{FowlerLiAnantuaFig5d
.png}{0.47\textwidth}{(d)}
         }
\caption{Showing evidence for a short-circuit current in the disk/coronal region of GRMHD simulations cited in the text. Panel (a) shows the vertical z-component of conical jet currents emerging up and down from the accretion disk and returning in a close-fitting jacket surrounding the jet 
(note $J_z$ goes down in the dark green regions and up in light green regions); the inner $-20M<x,z<20M$ region is expanded below, where $M=0.5R_S$ (Schwarzschild radius). Panel (c) expands the region in Panel (a) closest to black hole, with overlaid drawings following possible looping paths of short-circuit currents.  Panel (b) shows MRI-driven 3-D activity in the radial component of the current; the inner $-20M<x,z<20M$ region is expanded below in Panel (d) (note $J_x$ goes left in the dark green regions and right in light green regions).  \label{Fig5}}
\end{figure}

Future work might disclose other connections to our model, for example, the short-circuit $j_r$ current in the disk/corona in GRMHD simulations. Preliminary evidence of a short-circuit is shown in Figure \ref{Fig5} 
analyzing data from a GRMHD simulation in 
\citet{mckinney2012}. 
Shown are trajectories of $j^{1/3}$ (to add contrast) in Cartesian x-z planes cutting through 3D current structures 
($|x|$ being $r$ at $y = 0$). The predicted short-circuiting current loops  are indicated by overlays 
in Panels (c) and (d), occurring within a radius $r = a = 20M \equiv 10R_S $, 
this being the zone of greatest gravity-driven MRI activity as calculated in Section \ref{sec:jetpara}. 
The tilting geometry of actual loops gives evidence of looping both in 
$j_z$ in Panel (c) and in $j_x$ in Panel (d). 


The creation and propagation of magnetic towers (or pinches) is well understood from laboratory experiments on spheromaks \citep{hooper2012,zhai2014}. 
Spheromaks using plasma guns to replace the accretion disk produce sub-Alfvenic, 
collimated, blunt-nosed structures like Figure \ref{Fig1} that justify our calculation of $dL/dt$ 
independent of $r$ in Equation (5c) (see Paper II). 
This collimation persists the full length of the jet no matter how long the jet length grows. 
That jets produced by AGNs are also magnetically collimated over long distances is now supported by new radio telescopic observations \citep{giovannini2018} which corroborate earlier evidence from Faraday rotation measurements indicating the presence of collimated current channels with radii of order 10 $R_S$ (Schwarzschild radii) far from black holes \citep{owen1989,kronberg2011,lovelace2013}; and the corresponding evidence near the black hole \citep{zamaninasab2014,kim2018}. 
In Section \ref{sec:ele}  and Paper II, we show that contrary evidence based on conical synchrotron images represented by the dashed line in Figure \ref{Fig1} can be explained as a few wandering field lines that spread the radiated power but do not spoil the overall collimation of the current. 
Field lines wandering around magnetically-collimated jets have been observed in three-dimensional (3D) magnetic tower simulations, both non-relativistic \citep{nakamura2006} and relativistic \citep{guan2014}. 
The large dimensions of the nose agree with data on radio lobe dimensions, represented by the closed flux in Figure \ref{Fig1} \citep{diehl2008}. 
And evidence that this huge jet/lobe boundary pushes away the ambient as predicted is shown by bubbles seen in galaxy clusters \citep{mcnamara2007}. 

\subsection{Jet Parameters}
\label{sec:jetpara}
 
We conclude this Section with a review of numbers needed for our accelerator model in later sections. Following Paper III, we note that the key quantities needed concern the final jet velocity $\frac{dL}{dt}$ and numbers associated with the Central Column where most of the gravitational power is deposited. As in Papers I and II, the Central Column radius $a$ 
is defined as the inner radius of a Diffuse Pinch characterized by Keplerian rotation in the disk, 
while the Central Column at $r < a$ is described by the electric circuit in Equation (5b). 
The Diffuse Pinch helps guide and stabilize the Central Column flow, as discussed in Paper II.  
We obtain:
\begin{align}
IV &= \frac{a B_{a \phi}c}{2}\left(b\left(\frac{a \Omega_a}{c}\right)a \left|B_{az}\right|\right) = f\left(\frac{1}{4}\dot{M} c^2\right) \tag{8a} \\
I &= \int_0^a 2 \pi r \, dr \, j_z = \frac{a B_{a \phi} c}{2} \tag{8b} \\
V &= \int_0^a dr \, E_r = b\left(\frac{a \Omega_a}{c}\right)a \left|B_{az}\right| \tag{8c} \\
\frac{1}{2} \dot{M} \Omega_a &= a B_{a \phi} |B_{az}| \tag{8d}~~,
\end{align}
where Equation (8d) is the angular momentum conservation equation, Equation (6c), evaluated at $r = a$. 
Here the subscript $a$ denotes quantities at $r = a$, and $f$ is the efficiency of converting accretion 
power $\left(\frac{1}{4} \dot{M} c^2\right)$ into Central Column jet power $IV$, with current $I$ 
and voltage $V$ with fitting factor $b$. 
The accretion rate is taken to be $\dot{M} = M/\tau$ with typical jet lifetime $\tau = 10^8$ years 
\citep{colgate2004}; \citep{beskin2010}. 
Simultaneous solution of Equations (8a) and (8d) gives:
\begin{align}
\frac{a \Omega_a}{c} &= \left(\frac{f}{b}\right)^{1/2} = 0.2 \quad ; \quad \frac{a}{R_S} = \frac{b}{2f} = 10 \tag{9a} \\
b &= 5 \quad ; \quad	f = \frac{1}{4} \tag{9b} \\
a^2 B_{a \phi}|B_{az}|	&=	\frac{1}{2} \dot{M}c \left(\frac{f}{b}\right)^{1/2} \tag{9c} \\
B_{a \phi} &= |B_{az}| \equiv B_a = 1.5 \times 10^3 {M_8}^{-1/2} \text{ tower} \tag{9d} \\
(B_{a \phi})_{\text{cone}}  &=	E_r	= \frac{a \Omega_a}{c}|B_{az}| = \left(\frac{f}{b}\right)^{1/4}(B_{a \phi} )_{\text{tower}} \tag{9e} \\
n_I	&= \frac{I(a)}{e \langle v \rangle A} = \frac{ B_{\phi} rc}{2e \langle v \rangle A} \tag{9f} \\
V &= a B_a = 2\left(\frac{I}{c}\right) \tag{9g}~~.
\end{align}
Here $M_8$ is $M$ in units of $10^8$ solar masses; Keplerian $\Omega_a = \sqrt{\frac{GM}{a^3}}$ 
and $R_S = \frac{2GM}{c^2} = 3 \times 10^{13} M_8$; $f_{\text{dis}} = \frac{1}{4}$ (half from shocks, 
half from ion acceleration, as in Paper II); and we take $f = \frac{1}{4}$ from Paper I and $b = 5$ 
as discussed in Paper II, Appendix B. 
Equation (9d) is obtained from Equation (9c) using $B_{a \phi} = |B_{az}|$ for magnetic towers, 
from Paper I; to be compared with that when a conical jet is launched in Equation (9e). 
The ratio $(B_\mathrm{\phi\ cone}/B_\mathrm{\phi\ tower})=(f/b)^{1/4}$ accounts for current 
ratio $(I_\mathrm{\phi\ cone}/I_\mathrm{\phi\ tower})=0.5$ at $r=a$ in 
Figure \ref{Fig4}, using $I\propto B_\phi$ in Equation (8b). Equation (9g) applies
Equations (9a) and (9b) to Equations (8b) and (8c). Note that $a, \Omega_a,b ,f $ and the total jet power IV are
unchanged (except as $M_8$ changes) during the transition from a conical jet to a
magnetic tower. Note also the Diffuse Pinch current omitted here [($I(r)-I_A$) at $r > a$ 
in Figure \ref{Fig4}], though $40 \%$ of the total, only accounts for about $15\%$ of the total dynamo power. 

The other number we need for accelerator calculations is the density in the Central Column, given in Equation (9f), where $\langle v \rangle \rightarrow c$ and $A = \pi a^2$. 
As is discussed in some detail in Paper II, Appendix A, the fact that the magnetic tower jet is ejected vertically with negligible $B_r$ requires that an electrostatic sheath form to eject ions against gravity (for our sign convention).

Why the jet current $I$ becomes the constant value in Equation (8b) is due to macrostability of the system, as discussed in Paper II, based on earlier results in \citet{fowler2009}. 
That Equation (9f) is always correct, despite mass loading along the jet (not expected, Paper II) or pair-production, follows from current suppression by two-stream instability. How two-stream scattering of electrons (and positrons) kills their current in our reference frame is shown in Equation (C5). That the net ion density cannot exceed that in Equation (9f) follows from acceleration of all ions to speed c, requiring that radial hyper-resistive transport expel any excess of ions beyond $j_z=n_iec$ emerging from an electrostatic sheath in the disk corona.

Another feature of magnetic towers is their radial confinement by ambient pressure that finally limits the growth of $R$ even as $L$ continues to grow. Following \citet{lynden-bell2003}, the final $R$ and other jet parameters are given by:
\begin{align}
\frac{B_{\phi}(R)^2}{8\pi} &= p_\mathrm{amb} \tag{10a} \\
\frac{dL}{dt} &= \frac{cf_{\text{dis}}}{\ln\left(R/a\right)} \approx  0.01c \tag{10b} \\
L &\approx 0.01 c \tau = 10^{24}\text{ cm} \tag{10c} \\
R &= 0.1 L \tag{10d}
\end{align}
%
%
%
where again $\tau = 10^8$ years is the typical AGN jet lifetime; and $R$ in $\ln\left(\frac{R}{a}\right)$ is the radius of giant radiolobes bounded by the ambient pressure \citep{diehl2008}. 
Earlier in time, the ambient pressure in Equation (10a) is replaced by ram pressure due to jet inertia. 
Following Paper II, we take $\ln \left(\frac{R}{a}\right) \approx 20$ for typical radiolobe dimensions. 
The final jet length in Equation (10c) is consistent with observations \citep{krolik1999}.

\subsubsection{Importance of Large R and Current Loop Closure}

Our accelerator model will require that current loops begin to close at the nose, and that the nose radius $R$ be very large as predicted in Equation (10d). That current loops will try to close follows when slowing down gives $c^{-1}\mathbf{j}^*\times\mathbf{B}\approx \triangledown p_\mathrm{amb}$, equivalent to Equation (7b) when $dL/dt \ll c$. Then the procedure giving Equation (4a), applied to the nose using Equation (10a) to eliminate $p_\mathrm{amb}$, gives $I_\perp\approx I(r/R)^2$, showing that for large R there is essentially no short-circuit over most of the nose of a slowly-evolving jet. Absent a short-circuit at the nose, $\mathbf{j}\times\mathbf{B}=\mathbf{0}$ in the jet return so that, to access dynamo power $\mathbf{j}\cdot\mathbf{E}$ 
inside the disk, the return current must follow flux surfaces encircling an O-point inside the disk, located at $r=R_0$ where $B_z=B_r=0$. A quasi-static poloidal seed field necessarily has such an O-point which, however, may be inaccessible at speed c. Even so, Equations (3a) and (3b) giving simultaneous growth of $A_r\approx |A_\phi|$ exceeding the seed field in the disk would create a peak in $\psi=rA_\phi$ near $r=a$ 
giving an O-point where $B_z=r^{-1}\partial \psi/\partial r = 0$, growing eventually to $R_0 = 0.001L$ as derived in Paper II. 

\section{ION ACCELERATION IN AGN JETS}
\label{sec:ionacc}

We now begin our discussion of how magnetic energy is converted to UHE cosmic rays, 
using known physics of plasma turbulence discussed in Appendix B. 
For simplicity, we assume ions to be protons, known to be constituents of the most 
energetic UHECRs \citep{cronin1999}, though the model applies to any ion species. 
As noted in Section \ref{sec:ele
}, we neglect positron production. 
The model also describes electron acceleration yielding synchrotron radiation, discussed in Section \ref{sec:ele}  and Appendix C. 

We begin with two points distinguishing our model of UHE cosmic rays. 
First, while AGN jets are mainly observed by synchrotron radiation, the synchrotron power is known to 
be a negligible fraction of overall AGN luminosity and electrons play a secondary role in our model of 
ion acceleration. 
Second, 
as discussed in our Paper III, our cosmic ray accelerator model is a two-stage ion accelerator. 
The first stage occurs in the Central Column, limited by ion radiation to energies well below UHECRs. 
The second stage occurs in the nose-end of the jet. In both stages, acceleration is due to plasma 
turbulence (hyper-resistivity in Ohm's Law). What distinguishes these two regions is the size of the ion 
Larmor radius $r_{Li}$. In the Central Column, $r_{Li} \ll a$ giving MHD current-driven kink modes 
as the main source of hyper-resistivity. In the nose with magnetic flux width $\Delta$, 
the fall of $B_\phi \propto 1/r$ finally yields $r_{Li} \approx \Delta$ whereupon ions 
resonating with electron drift waves are known to produce a powerful turbulence (specifically the
DCLC instability explained in Section 3.3.2) 
that could both accelerate ions to $10^{20}$ eV energies and eject enough ions as cosmic rays 
to account for the UHECR intensity on Earth (see Section \ref{sec:CRs}).
 
The electrons behave passively in DCLC turbulence so they are not accelerated in the nose, 
while ions resonant with the electron waves in the nose are strongly accelerated. For a different reason, 
electrons play a passive role in kink instability in the Central Column. Again this is due to kinetic 
effects not included in MHD; namely, the two-stream instability between counter-flowing ions and 
electrons accelerated by the kink mode turbulence. Because of their difference in rest mass, 
two-stream instability scatters electrons sufficiently to eliminate the electron current in our reference frame, 
while ions are not much affected. Thus the kink mode accelerator in the Central Column produces 
current as a mono-energetic ion beam with sufficient energy to produce DCLC instability in the nose. 

The mathematics supporting the above picture is developed below. A key feature is the role 
of turbulence-driven hyper-resistivity in Ohm's Law. Like earlier  cosmic ray models, hyper-resistivity 
depends on correlations in turbulent fluctuations to give additive additions to particle energy. 
Self-correlated hyper-resistive ion acceleration by kink modes was observed in the SPHEX 
experiment, discussed below (see Paper III). Two-stream instability is discussed in Appendix C.2.

\subsection{Acceleration by Turbulence}

Our accelerator model begins with the relativistic acceleration equation, with acceleration of 
momentum $\mathbf{p}$ of a particle given by:
\begin{equation}
\frac{d\mathbf{p}}{dt}		= 	e\left(\mathbf{E} + \frac{1}{c}\frac{\mathbf{p}}{m \gamma_{L}} \times \mathbf{B} - \mathbf{E}_{\text{rad}}\right) \tag{11}
\end{equation}
where $\mathbf{E}_{\text{rad}}$ represents radiation loss. 
While taking $\mathbf{p} \cdot \frac{d\mathbf{p}}{dt}= e \mathbf{p} \cdot \left(\mathbf{E} - \mathbf{E}_{\text{rad}}\right)$ confirms that only $\mathbf{E}$ can accelerate ions parallel to $\mathbf{B}$ in our reference frame fixed in the disk, by now a number of magnetic acceleration mechanisms have been identified in which $\mathbf{B}$ invokes $\mathbf{E}$ in moving structures (shocks, clouds, etc.). 
Acceleration by turbulence can be thought of this way, whereby accelerated particles 
encounter fluctuating electric fields in a self-organized way in a plasma. 

Formally, in our model energy conversion comes from $\int d\mathbf{x} \langle \mathbf{j} \cdot \mathbf{E} \rangle \equiv f_{\text{conv}}IV$ where the integral is over the entire structure of the magnetic jet and $\langle ... \rangle$ represents an average over the toroidal angle $\phi$ and an average over fluctuations. 
The integral is dominated by the Central Column which carries most of the power to the nose in Figure \ref{Fig1}. The average $\langle ... \rangle$ yields a turbulence-generated $\langle \mathbf{E} \rangle$ that could serve as a quasi-steady accelerator. 
Most of the kinetic power is contained in $\langle \mathbf{j} \rangle \cdot \langle \mathbf{E} \rangle$, accounting for the observation of quasi-steady acceleration of ions in the SPHEX spheromak mentioned above. 
In this Section we will focus on the quasi-steady acceleration of ions, and defer electron acceleration to Section \ref{sec:ele}. 
To calculate ion acceleration, we determine $\mathbf{E}$ from the relativistic form of Ohm's Law derived from Equation (1a), now retaining the relativistic correction and the Hall term.
We drop terms on the left hand side to obtain: 
\begin{equation}
\mathbf{E} - \sum \left[ \int d\mathbf{p} \left(\frac{f}{n}\right)\left(\frac{m_e \gamma_{Le}}{m \gamma_L}\right)\frac{1}{c^2}\mathbf{u}\left(\mathbf{u} \cdot \mathbf{E}\right)\right] + \frac{1}{c}\mathbf{v} \times \mathbf{B} - \left(\frac{\mathbf{j}}{n e c}\right) \times \mathbf{B} - \mathbf{D} = \mathbf{0} \tag{12}
\end{equation}
where $\mathbf{u} = \left(\mathbf{p}/m \gamma_L\right)$.
The Hall term applies only to $\mathbf{E}_{\perp}$ giving MHD. From the radiation discussion below, we will learn that always the ion Lorentz factor $\gamma_{Li} \geq \gamma_{Le}$, which allows us to order all ion terms in Ohm's Law as the ratio of electron rest mass to ion rest mass, hence negligible. 
Dropping the Hall term, we obtain for relativistic electrons:
\begin{align}
\mathbf{E}_{\perp} + \frac{1}{c}\mathbf{v} \times \mathbf{B} &= \mathbf{D}_{\perp}	\tag{13a} \\
C_{\text{Rel}}\mathbf{E}_{\parallel} & \equiv \mathbf{E}_{\parallel} - \sum \left[ \int d\mathbf{p} \left(\frac{f}{n}\right)\left(\frac{m_e \gamma_{Le}}{m \gamma_L}\right)\frac{1}{c^2}\mathbf{u_\parallel}\left(\mathbf{u} \cdot \mathbf{E}\right)\right] 
= \mathbf{D}_{\parallel} \tag{13b} \\
C_{\text{Rel}}	&\approx	1-\int d\mathbf{p}f_{0e}\left[(u_{||}^2/c^2)+u_{||}(\mathbf{u}_\perp\cdot\mathbf{E}/E_{||})\right] \approx \int d\mathbf{p}f_{0e}(u_\perp^2/c^2) \approx 1 \tag{13c}
\\
\mathbf{D}	&=	- \frac{1}{c}\langle \mathbf{v}_1 \times \mathbf{B}_1 \rangle \tag{13d}
\end{align}
where we use $(u_{||}^2+u_\perp^2)=c^2$ and $\int d\mathbf{p}f_{0e}u_{||}=0$ by two-stream instability discussed in Section C.2. Here, we write out $\mathbf{D}$ explicitly giving hyper-resistivity for MHD fluctuations. 
The bracket $\langle ... \rangle$ indicates an axisymmetric smooth average over fluctuations giving Equation (3a). Equation (13a) yields MHD jet propagation in Section 2 while $C_{\text{Rel}}\mathbf{E}_{\parallel} = \mathbf{D}_{\parallel}$ is the relativistic accelerator in the Central Column of the jet. 
Note the crucial role of two-stream instability in defeating a relativistic cancellation of the accelerator parallel to $\mathbf{B}$ in our reference frame.

\subsection{Ion Acceleration in the Central Column}
	
We make the assumption, justified later, that ions and electrons in the Central Column can be described by orbits consisting of circles with small Larmor radii $\ll a$, gyrating around a ``guiding center'' momentum $p_{\parallel}$ mainly directed along magnetic field lines. Using Ohm's Law, the acceleration equation, Equation (11), becomes:
\begin{align}
\frac{dp_{\parallel}}{dt} &= e \left(E_{\parallel}  	-  E_{\text{RAD}}\right) \tag{14a} \\
E_{\parallel} &= D_{\parallel} =  -\langle \frac{1}{c}\mathbf{v}_1 \times \mathbf{B}_1 \rangle_{\parallel} =  - \langle \frac{\left(\mathbf{E}_1  \times \mathbf{B}\right) \times \mathbf{B}_1}{B^2}\rangle_{\parallel} \tag{14b} ~~,
\end{align}
where $\mathbf{E}_1$, $\mathbf{B}_1$ and $\mathbf{v}_1$ are 3D fluctuations around the mean fields due to kink modes driven by the jet current, and for MHD we drop the term $f_1 \mathbf{E}_1$ in Equation (13c), leaving $\mathbf{v}_1 \times \mathbf{B}_1$ with $\mathbf{v}_1 = \int d\mathbf{p} (\frac{f_1 \mathbf{u}}{n})$.
Equation (14b) can be rewritten as:
\begin{equation}
\mathbf{E}_{\parallel} =	- \langle \mathbf{E}_{1 \perp} \cdot \mathbf{B}_{1 \perp} \rangle \left(\frac{\mathbf{B}}{B^2}\right) \tag{15}~~.
\end{equation}

The main issue is whether actual 3D fluctuations correlate to produce a finite $E_{\parallel}$, especially for ideal kink modes argued to dominate behavior in the Central Column of astrophysical jets in Paper II. 
That Equation (15) with ideal MHD perturbations does produce a finite mean electric 
field was demonstrated by careful measurement during kink-mode instability in the SPHEX 
experiment under conditions when resistivity was negligible \citep{al-karkhy1993}. 
Acceleration of ions by this electric field has been verified directly \citep{rusbridge1997}. 

Theoretical evidence that ideal MHD fluctuations can correlate is shown in 
Figure 6 giving just the inductive contribution to $E_{\parallel}$ produced by ideal MHD kink modes \citep{mcclenaghan2014}, using the non-linear non-relativistic GTC PIC code \citep{deng2012}. 

\begin{figure}[H]\nonumber
\begin{align}
  \hspace{0cm} \includegraphics[height=200pt,trim = 6mm 1mm 0mm 1mm]{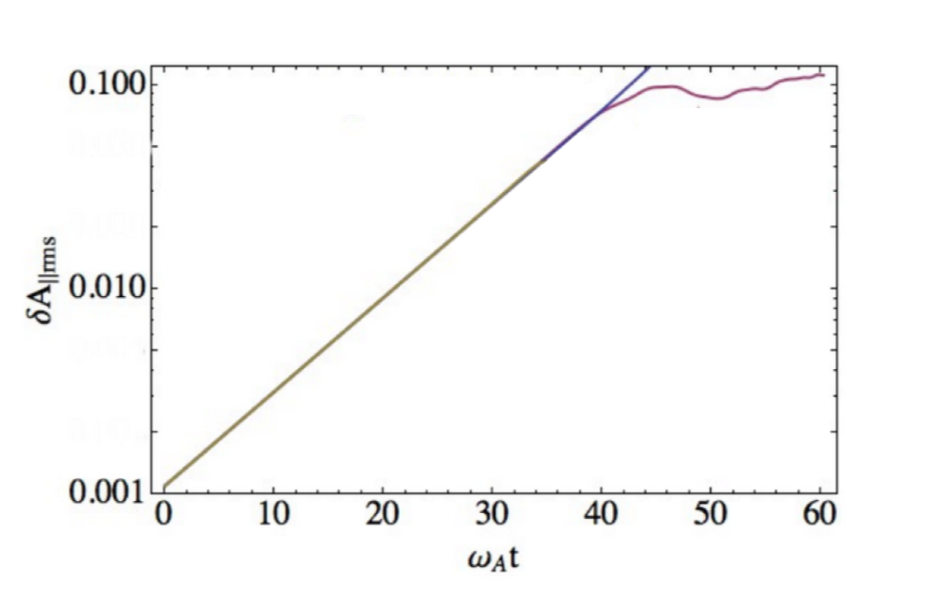} & 
\end{align}
\caption{Showing the growth and saturation of the inductive component of the hyper-resistive electric field parallel to the magnetic field due to kink mode turbulence in the jet Central Column in Figure \ref{Fig1}.}\label{Fig7PoloidalRzFlux}
\end{figure}

\subsubsection{Hyper-Resistivity as Diffusion}

Given favorable correlations, we estimate the magnitude of $E_{\parallel}$ as follows. We write $E_{\parallel}$ as:
\begin{equation}
E_{\parallel} = - \frac{1}{c}\langle \mathbf{v}_1 \times \mathbf{B}_1  \rangle_{||}  = \eta_H j_{\parallel} \approx  \left(\frac{D^H_{r}}{ac}\right)B_a \approx  \left(\frac{a}{ct}\right)B_a \approx \frac{V}{ct} \tag{16}~~,
\end{equation}
using $j_{\parallel} \approx \frac{c B_a}{4 \pi a}$ and $a B_a \approx V$ by Equation (9g). 
Equation (16) represents magnetic relaxation by hyper-resistive current diffusion \citep{fowler2007}, 
with $D_{||}$ in Equation (13b) giving 
$\eta_H = D_{||} / j_{||} = - \langle \frac{1}{c}\mathbf{v}_1 \times 
\mathbf{B}_1 \rangle_\parallel / j_{\parallel}$ symmetrically averaged over fluctuations $\mathbf{v}_1$ and $\mathbf{B}_1$. 
Note that $\eta_H$ has units of resistivity while $D_{\parallel}$ (which we are calling hyper-resistivity) has units of
the electric field. The term with $D^{\ H}_{r} = (c^2\eta_H/4\pi)$ relates hyper-resistivity to diffusion of the current, 
by analogy with classical diffusion via ordinary collisional resistivity. This characteristic relationship of 
hyper-resistivity in any direction $\mathbf{x}$ producing diffusion perpendicular to $\mathbf{x}$ 
will recur in considering ion acceleration by diffusion in Equation (21b).

We expect fluctuation amplitudes determining $D^H_{r}$ to saturate at levels not exceeding those 
required to flatten the known stability parameter $\lambda = \frac{4\pi}{c}\left(\frac{j_z}{B_z}\right)$ 
over a radius $R_1 \approx a$ where internal kink modes are localized, giving then the local 
source of free energy driving fluctuations with little effect on the current profile at $r > R_1$ \citep{fowler1968}. 
To extrapolate to astrophysical dimensions, while avoiding the need to calculate correlated fluctuations in detail, we approximate $D^H_{r} t  = R_1^2 \approx a^2$, giving $D^H_{r} \approx \frac{a^2}{t}$, where $t$ is the time required to reach quasi-steady state.
Taking $D^H_{r} \approx \frac{a^2}{t}$ gives the result for the accelerating electric field on the far right hand side of Equation (16). 

\subsubsection{Voltage Drop in the Central Column, Curvature Radiation}

We apply Equation (16) to $E_{\parallel}$  averaged over the duration of the jet, giving, at any elapsed time $t$, $L = t(\frac{dL}{dt}) = 0.01 ct$ from Equation (10b). 
Using also Equation (16), we obtain: 
\begin{equation}
\Delta V = 	LE_{\parallel}  \approx	0.01 V \tag{17}
\end{equation}

That this $\Delta V$ is independent of the elapsed time $t$ is a result of localization of internal kink modes to $k_z a \approx 1$ mentioned above. 
Then, even if the disk current profile itself does not relax, within a few wavelengths the local effect of internal kink modes can cause the jet current profiles to relax toward a stable state. 
As the system evolves non-linearly, instability growth rates and $D^H_{r}$ diminish as the current density approaches this stable profile. 
Then, the longer the duration, the weaker is the time-averaged $D^H_{r}$ needed to spread the current toward a stable profile. 
Implicit in this argument is the slow evolution of the $\lambda(r)$ profile driving kink instability, on the $\dot{M}$ timescale $\tau$, which is also the lifetime of the jet.

The $\Delta V$ in Equation (17) represents dissipation in the Central Column that depletes the power available to accelerate ions to UHECR energies, but only slightly because of the large inductance giving the slow evolution of the jet length, $\frac{dL}{dt} = 0.01 c$. 
The dissipation is in the form of acceleration of electrons giving the correct power for observed synchrotron radiation, in Section \ref{sec:ele}, and ion acceleration that is also mostly dissipated as radiation due to charged particles following the curvature of twisted magnetic field lines. 
By Equation (14a), the maximum energy allowed by curvature radiation is given by $E_{\parallel} = E_{\text{RAD}}$ with 
\citep[][Equation (14.31)]{jackson1998}.
\begin{equation}
E_{\text{RAD}} = \frac{2}{3} e \alpha_C^2 \left(\frac{\beta_L \gamma_L}{c}\right)^4	\tag{18}
\end{equation}
Here the right hand side gives the radiation with $\alpha_C = \frac{c^2}{R_C}$ for relativistic ions or electrons following field lines with curvature radius $R_C$ for which the Lienard factor $\beta_L = 1$. The maximum Lorentz factor $\gamma_L$ allowed by $E_{\parallel} = E_{\text{curv}}$ is independent of particle mass, given by: 
\begin{align}
\gamma_L &\leq \left( \frac{3E_{\parallel} R_C^2}{2e} \right)^{1/4} \tag{19a} \\
\gamma_{CC} &= 	3.4 \times 10^7 {M_8}^{5/8} \tag{19b}
\end{align}
To get $\gamma_{CC}$, we have used Equation (18) with 
$E_{\parallel} = 0.01V/L$ by Equation (17), line curvature $R_C = a$, and all the values from Section \ref{sec:jetpara}.
A few ions at $R_C=r<a$ have $\gamma_\mathrm{L}<\gamma_\mathrm{CC}$. Note that $\gamma_{CC}$  representing the maximum acceleration energy in the Central Column is far below $\gamma_L \approx 10^{11}$ ($10^{20}\text{ eV}$) required to explain UHECRs, hence the need for additional acceleration in the nose. 
On the other hand, that the Central Column is otherwise an excellent accelerator follows from a calculation of the ion Larmor radius $r_L = v_{\perp} /\omega_{c i}$ and pressure parameter $\beta_{CC}$ in the  Central Column, giving:
\begin{align}
\beta_{CC} 	&= \frac{8 \pi n_i m_i \gamma_{L}c^2}{B_a^2} = \frac{4r_{L 0}}{a} \ll 1 \tag{20a} \\
\left(\frac{r_{L 0}}{a}\right)_{CC}	&\equiv \frac{m_i \gamma_{CC} c^2}{a e B_a} = 2.5 \times 10^{-4} {M_8}^{1/8} \ll 1	\tag{20b} \\
\left(r_L\right)_{CC}	&= \frac{v_{\perp}}{\omega_{c i}} = r_{L 0}\left(\frac{v_{\perp}}{c}\right)  \tag{20c}
\end{align}
where in Equation (20c) $\mathbf{v}_{\perp}$ is the ion velocity.  
In Equation (20a) the numerator is the ion energy and we use the density 
$n_i  = n_I  = \frac{I}{e \langle v \rangle A}$ in Equation (9f) with 
$\langle v \rangle \approx c$, $A = \pi a^2$ and $I$ in Equation (8b), 
and in Equation (20b), we use $\omega_{c i} = \frac{eB}{m_i\gamma_L c}$ 
with $B = B_a$ at $r = a$ and $\gamma_L = \gamma_{CC}$ in Equation (19b). 
We see that the low beta giving strong magnetic collimation is 
coincident with small $\frac{r_L}{a}$ giving good ion confinement.

That ions in the Central Column should run away to the energy in Equation (19b) and 
that these runaway ions carry most of the current is shown as follows. 
Briefly, for $\Delta V$ from Equation (17) with numbers from Section \ref{sec:jetpara}, 
both ions and electrons reach a maximum $\gamma_L = \gamma_{CC}$ at a distance 
$0.02\text{ L}$. But two-stream instability between the oppositely-accelerated 
relativistic ions and electrons spreads electron momentum between 
$- \gamma_{CC} m_e c < p_{\parallel e} < +  \gamma_{CC}m_e c$. 
As shown in Appendix C.2, this kills the electron current, while by momentum 
conservation the corresponding ion momentum spread is less by a factor 
$\frac{m_e}{m_i}$ so that the ion distribution is approximately a delta-function 
around $p_{\parallel i} = \gamma_{CC} m_i  c$ carrying current at speed $c$. 
Details of two-stream instability are discussed in Appendix C.

\subsection{Ion Acceleration in the Nose}

Internal kink instability due to $j_z$ occurring in the Central Column, as analyzed in Paper II 
and used in Section 3.2, should persist for some distance into the nose driven by $j_r$ 
as the current turns radially in the nose, now less affected by curvature radiation as 
$B_{\phi}$ falls like $1/r$ giving curvature parameter $\alpha_C \approx 1/r$. 
But a new acceleration mechanism arises due to conditions in the nose region 
different from those in the cylindrical Central Column. 
One new condition concerns the sensitivity of the curvature radiation, Equation (19a), to $R_C$ 
that allows ions to be accelerated to the full voltage $V$ in the nose region where $R_C = r$ becomes large. 
Another new condition concerns the escape of ions from the nose, escape itself 
turning out to be a necessary ingredient for the powerful acceleration mechanism 
not present in the Central Column. 

There are two ways that ions can escape from the nose. 
First, while ion orbits are magnetically confined in the Central Column, in the absence of 
an electric field ion orbits in the nose are not confined due to drifts of guiding centers in $z$ 
due to curvature of field lines and gradients in the magnetic field intensity.  
Opposite drifts by electrons establish an electric charge giving $\mathbf{E} \propto T_e$, 
the electron temperature, which can cause ions to escape by $\mathbf{E} \times \mathbf{B}$ motion. 
In addition, ions can also escape by hyper-resistive diffusion.

\subsubsection{Hyper-Resistive Diffusion in the Nose}

As in Section 3.2.1, we represent hyper-resistivity in the nose as diffusion, noting that 
an electric field larger than that produced by charge separation is created if turbulence in 
the nose causes ions to escape by diffusion. To see this, consider the radial component 
of the relativistic Ohm's Law in Equations (13a) and (13b). 
In the nose, $\mathbf{j} \times \mathbf{B} = \mathbf{\underline{z}}(j_r B_{\phi} - j_{\phi} B_r)$ with unit vector $\mathbf{\underline{z}}$, and electron flow $\mathbf{v}_e$ is either that due to drifts in the $z$-direction or that parallel to $\mathbf{B}$ giving $v_{e r} < \left(\frac{B_r}{B}\right) c$ so that we can drop the relativistic correction to $E_r$. 
Applying Equation (13a) in the $r$-direction (perpendicular to $\mathbf{B}$) gives:
\begin{align}
E_r	&= - \frac{1}{c}\frac{\p A_r}{\p t} -\frac{\p \Phi}{\p r} = \frac{1}{c} v_z B_{\phi} + D_r \tag{21a} \\	
\left(E_r\right)_{\text{accl}} &= -\frac{\p \Phi}{\p r}	=  \frac{1}{c}\left(\langle v_z \rangle  - \frac{dL}{dt}\right)B_{\phi}	+ \frac{\langle D^H_{z} \rangle}{c \Delta}B_{\phi} \tag{21b} \\
\int_a^R dr \, \left(E_r\right)_{\text{accl}} &= f_{\text{conv}} \left(V - \Delta V\right) \approx f_{\text{conv}} V \tag{21c}			
\end{align}
Equation (21b) giving the cosmic ray accelerator $\left(E_r\right)_{\text{accl}}$ is obtained as follows. 
Early in jet propagation when $\mathbf{D} = \mathbf{0}$ in the jet and nose, all of the voltage drop in the jet (giving net zero voltage around the loop) can be approximated as $- \frac{1}{c}\frac{\p A_r}{\p t} = \frac{1}{c} v_z  B_{\phi} =         
(\frac{1}{c}\frac{dL}{dt})B_{\phi}$ which appears as an inductive electric field in our disk-centered reference frame. 
It is this inductive field that serves to build up $B_{\phi}$ behind the nose advancing at velocity $\frac{dL}{dt}$. 
Even as shocks and cosmic ray acceleration slow down $\frac{dL}{dt}$ by Equation (5b), 
it continues to be true that it is $- \frac{1}{c}\frac{\p A_z}{\p t} = (\frac{1}{c}\frac{dL}{dt})B_{\phi}$ that 
builds up $B_{\phi}$, giving Equation (21b) with $\left(-\frac{\p \Phi}{\p r}\right)$ in our reference frame 
serving as the cosmic ray accelerator in the nose. The final step relates the radial hyper-resistivity 
$D_r$ to vertical hyper-resistive diffusion $\langle D^H_{z}  \rangle$  that ejects ions in the $z$-direction, 
analogous to radial kink mode diffusion producing $E_{\parallel}$ in Section 3.2.1. 
Here $\langle D^H_{z}  \rangle$  is averaged over the energies of escaping ions, 
and $\Delta$ is the flux width 
in the nose. 
The disk-averaged vertical velocity $\langle v_z \rangle$  is dominated by disk ions giving 
$\langle v_z \rangle \approx dL/dt$ and $(E_r)_{accl} \approx (\langle D^H_{z} \rangle /c\Delta)B_{\phi}$ , which shows that it is the escape of cosmic rays vertically that drives cosmic ray acceleration radially (as weak  radial escape drives vertical acceleration in the Central Column).
The $\left(E_r\right)_{\text{accl}}$ in Equation (21b) adds to any residual $E_{\parallel}$ due to kink modes in the nose as field lines bend radially to create the nose, giving a continuous $B_{\phi}$ while $B_z$ turns into  $B_r$. 
Equation (21c) relates $\left(E_r\right)_{\text{accl}}$ in the nose to the kinetic conversion (acceleration) efficiency $f_{\text{conv}}$ in Equation (5e), aside from $\Delta V \approx 
0.01 \, V$ due to kink acceleration, in Equation (17), which we neglect here. 

$\qquad$
\newline

\subsubsection{DCLC Instability}           

The origin of the cosmic ray accelerator field $\left(E_r\right)_{\text{accl}}$ is not yet specified. 
Strong diffusion giving a large $\left(E_r\right)_{\text{accl}}$ would inevitably arise due to the 
Drift Cyclotron Loss Cone (DCLC) instability \citep{post1966}, 
not included in MHD but known from fusion 
research and characteristic of non-Maxwellian ion distributions such as the runaway ion beam 
created by kink mode acceleration in the Central Column. 
This runaway beam injecting energetic ions into the nose serves as the first stage of cosmic 
ray acceleration, DCLC in the nose being the second stage. 
Both the two-stream instability mentioned at the end of Section 3.2 and the 
DCLC instability are caused by ion excitation of electrostatic electron waves 
$\propto \exp \left\{ i(\mathbf{k} \cdot \mathbf{x} - \omega t)\right\}$ satisfying 
the following dispersion relation derived in Appendix B, applicable to DCLC as discussed in Appendix B.2. 
We obtain: 
\begin{align}
1-F_i(\omega,\mathbf{k}) &= \left(\frac{{k_{\parallel}}^2}{k^2}\right) \left(\frac{{\omega_{pe}^*}^2}{\omega^2} \right)+ \left(\frac{{k_{\perp}}^2}{k^2}\right) \left(\frac{\varepsilon}{k}\right) \left(\frac{{\omega_{pe}}^2}{\omega_{ce} \omega}\right) \tag{22a} \\
	               		&\rightarrow \frac{r_{L i}}{\Delta}\left(\frac{1}{k_{\perp} r_{L i}}\right)\left(\frac{{\omega_{pe}}^2}{\omega_{ce}\omega}\right)	\qquad\text{at nose} \tag{22b}
\end{align}
where $\omega_{pe}^* = (4\pi ne / m_e \gamma^3_{Le})^{1/2}$ and $\omega^2_{pe}/\omega_{ce}\omega$ is independent of
electron mass (see Appendix B and C); $k_{\parallel}$ and $k_{\perp}$ are components parallel and 
perpendicular to $\mathbf{B}$; and $F_i \left(\omega, \mathbf{k}\right)$ is the ion term derived in Appendix B. 
In the Central Column, initially the dominant electron waves are plasma oscillations with $k_{\perp}= 0$ on the right hand side. 
This is the two-stream instability that we found not to affect ions very much, in Section 3.2. 
But as ions enter the nose, new conditions allow the ions to excite electron ``drift waves'' giving DCLC instability resonant at the ion cyclotron frequency. 
Equation (22b) describes drift waves, with $k_{\perp} \gg k_{\parallel}$. 
Drift waves are caused by the electron density gradient given in the nose by 
$\varepsilon \equiv \frac{dn}{dz}/n = \frac{1}{\Delta}$. Why and when drift waves produce 
DCLC instability can be understood as follows. 
 
There are two conditions yielding DCLC instability. First, DCLC is a self-driven cyclotron in which the 
oscillating drift wave acts as the cyclotron accelerator but the energy comes from the ions themselves. 
It is for this reason that the drift wave frequency must resonate with ion rotation. The second, more important, 
condition concerns when ions can transfer energy to the drift wave. This aspect of DCLC physics 
is well-known from mirror fusion devices, where DCLC is the remaining 
electrostatic instability when Landau damping stabilizes all modes with finite $k_{\parallel}$ along $\mathbf{B}$. 
The DCLC with $k_{\parallel}= 0$ requires that the distribution function averaged over $p_{\parallel}$ be 
non-Maxwellian (Paper III), as is the case in mirror devices due to the ejection of low energy ions by 
an electrostatic potential needed to confine the electrons \citep{fowler1981}. 
The potential creates a hole in momentum space around $p_{\perp} = 0$, yielding free energy 
producing DCLC turbulence. An analogous situation occurs when ions accelerated parallel to 
$\textbf{B}$ in the Central Column encounter the nose. The DCLC instability is a mode with 
$k_{\parallel} = 0$ driven by a hole in the averaged distribution $f(p_{\perp}) = \int dp_{\parallel} f$  (Paper III).
As we show below, ions following bending field lines into the nose experience a centrifugal 
force that causes them to acquire a minimum $p_{\perp} = (r_{L 0}/a)p_{\parallel}$, giving then 
a hole in $f(p_{\perp})$ around $p_{\perp} = 0$ that can drive DCLC.

The DCLC instability occurs for a sufficiently large value of $\frac{r_{L i}}{\Delta}$, found by 
choosing $k_{\perp}$ and $\omega$ to minimize the value required to satisfy Equation (22b) 
for real $\omega$. 
The derivation in the relativistic limit is reviewed in Appendix B.2, giving the same criterion 
for the onset of instability as the well-documented result for the non-relativistic case if we simply 
replace rest masses by relativistic masses, giving a relativistic ion plasma frequency 
$\omega_{pi}= \left(\frac{4 \pi n e^2}{m_i\gamma_{L i}}\right)^{1/2}$  and a relativistic ion 
cyclotron frequency $\omega_{c i} = \frac{e B_{\phi}}{m_i\gamma_{L_i}c}$. 
Then the DCLC instability occurs if $r_L/\Delta > 0.4\left(\frac{\omega_{c i}^2}{\omega_{pi}^2}\right)^{2/3}$ 
\citep{fowler1981,post1966}. 

We rewrite the DCLC instability condition by eliminating $\omega_{c i}$ using $r_L = v_{\perp}/\omega_{c i} = \frac{m_i \gamma_i c^2}{e B_{\phi}}\left(\frac{v_{\perp}}{c}\right)$ with dominant field $B_{\phi}  = 2I/r$, and by eliminating $\omega_{pi}$ using $n$ from Equation (9f), giving for relativistic current carriers $n = \frac{I}{Aec} = \frac{r B_{\phi}}{2 e c A}$ with $A = 2 \pi r \Delta$ for return flux width $\Delta$.
Substituting these quantities into the DCLC instability condition, we obtain after some algebra:
\begin{align}
r_L/\Delta &> 0.4 \left(\frac{{\omega_{ci}}^2}{{\omega_{pi}}^2}\right)^{2/3} =	0.4\left(\frac{{B_{\phi}}^2}{4 \pi n m_i\gamma_{Li} c^2}\right)^{2/3} \notag \\
		&= 0.4 \left(\frac{e c A B_{\phi}}{2\pi r m_i\gamma_{Li}c^2}\right)^{2/3} = 0.4 \left[(\Delta/r_L)(v_{\perp}/c)\right]^{2/3} \tag{23a} \\
\frac{v_{\perp}}{c} &= \left(\frac{r_{L0}}{a}\right)\left(\frac{\gamma_{Li}}{\gamma_{L0}}\right) \tag{23b} \\
\frac{r_L}{\Delta} &\geq 0.6\left(\frac{r_{L0}}{a}\right)^{2/5}	\text{DCLC onset in the nose} \tag{23c}
\end{align}
where we use $n$ from Equation (9f) with $\langle v \rangle \approx c$. Substituting Equation (23b) into (23a) gives Equation (23c), with $\gamma_{Li} = \gamma_{L0}\equiv\gamma_\mathrm{CC}$ as ions enter the nose.

Equation (23b), with $r_{L0}$ in Equation (20b), is derived from the perpendicular momentum equation $\frac{dp_{\perp}}{dt} = \left(\text{Force}\right)_{\perp}$, as follows. 
Since the two-stream instability does not scatter ions very much, the orbital spin velocity $v_{\perp} = (p_{\perp} /m_i \gamma_L )$ in Equation (23b) is determined by balancing the centrifugal force, $m \gamma_L c^2 /r$, due to magnetic curvature radius $r$, against the restraining magnetic force, $e(v_{\perp}/c)B_{\phi}$ with $B_{\phi} = B_a \left(a/r\right)$.
This ignores the ion synchrotron radiative force, which is much weaker, so that $v_{\perp}$ in Equation (23b) is always maintained. 
This is the minimum $v_{\perp}$ required for ion confinement in a twisting magnetic field, giving then the hole in momentum space causing DCLC instability to occur.

\subsubsection{Onset of DCLC Instability}

Initially, the low $\beta_{\perp} \ll 1$ carried forward from the Central Column preserves a 
force free field as ions enter the nose, giving $B_z  \rightarrow B_r \approx B_{\phi}$ as field lines bend 
radially in the nose with flux width $\Delta \approx a/2$, obtained from force balance 
$j_r B_\phi = j_\phi B_r$, which gives $B_\phi/B_r = 2\Delta/a = 1$ using $B_\phi = B_a (a/r)$ and 
flux conservation $B_r = \pi a^2 B_a/ (2\pi r \Delta)$.  
Also, initially the DCLC instability condition is not satisfied at the entry to the nose. 
However, for parameters above and numbers from Section \ref{sec:jetpara}, 
even with no further acceleration by kink modes, the condition for DCLC instability would 
already be satisfied at a radius $r = R_{ac}$ where the DCLC condition in Equation (23c) is first satisfied. 

We calculate the radius $R_{ac}$ where DCLC commences as follows. We take $R_{ac}$ to be 
$r$ at the margin of DCLC instability, satisfying Equation (23c) with the equality sign with 
$r = R_{ac}$ and $r_L$ in the nose. 
We obtain:
\begin{align}
r_L	&= r_{L 0}\left(\frac{v_{\perp}}{c}\right)\left(\frac{r}{a}\right) = \left(\frac{r_{L 0}}{a}\right)^2 r \text{~~~~at  nose} \tag{24a} \\
r_L/\Delta  &= \frac{2r_L}{a} = 0.6\left(\frac{r_{L 0}}{a}\right)^{2/5} \tag{24b} \\
R_{ac}/a &=	(0.6/2)(a/ r_{L 0})^{8/5} = 1.7 \times 10^5 \,{M_8}^{-1/5} \tag{24c} \\
r_L/\Delta &= 2r_L/a = 0.02  {M_8}^{1/20} \tag{24d}
\end{align}
Again, $M_8$ is $M$ in units of $10^8$ solar masses. Equation (24a) uses Equation (23b), 
and Equation (24b) is Equation (23c) with the equality and $\Delta = a/2$. Equation (24c) comes 
from Equation (24b) using Equation (24a) with  
$r = R_{ac}$. Numerical values use Equation (20b). 
These results follow from the fact that, even though as ions flow into the nose scaling 
gives constant $v_{\perp}$ at its value in the Central Column until further acceleration occurs, 
the cyclotron frequency $\omega_{ci} \propto B_{\phi} \propto 1/r$ giving $r_{L i} \propto r$ that 
must eventually exceed a fixed $\Delta$ in the DCLC instability condition, Equation (23c).

\subsubsection{Ion Acceleration to UHECR Energies}

After the onset of DCLC instability at $r = R_{ac}$, two things happen. First, acceleration causes 
the ion energy $E$ to begin to increase. 
Secondly, DCLC scattering of ions gives $v_{\perp} \rightarrow c$ in about one cyclotron period, too fast 
for ion synchrotron radiation to prevent this as it did in the Central Column. 
Then, using $V = a B_a$ from Equation (9g), the relativistic ion Larmor radius in the nose 
becomes $r_L = \left(c/\omega_{ci}\right) = \left(\frac{r}{a B_a}\right)\left(\frac{m_i \gamma_i c^2}{e}\right)= 
r(E/eV)$ with an energy distribution f(E) due to downward scattering by DCLC. 
(Notation: note that here eV means multiply the accelerator voltage V by the electron charge, 
not to be confused with eV as a unit of energy.)

At the onset of DCLC, Equation (24d) shows that ions are still well confined within the initial flux 
width of the force free field. 
As the energy $E$ increases by DCLC acceleration, ions would remain confined only if the flux width 
widens to satisfy $\Delta > r_L$ for all confined ions. 
The flux width also determines a pressure parameter $\beta_{\perp}$ giving the vertical 
pressure balance. 
We calculate $\beta_{\perp}$in three steps: 
\begin{align}
\Delta &> r_L & &= [m_i c^2 \gamma_L / e B_{\phi}(a)(a/r)] = r (E/eV) \tag{25a} \\
\frac{dp}{dz} &= \frac{p}{\Delta} & &\approx -\frac{1}{c}\left[j_r B_{\phi} - j_{\phi} B_r\right] \approx - \left(\frac{1}{8\pi \Delta} \right)\left[{B_{\phi}}^2 - {B_r}^2\right] \tag{25b} \\
\beta_{\perp} &= (8\pi p / {B_{\phi}}^2) & &= 1  -  ({B_r}^2/{B_{\phi}}^2)	=	1 - \left(\frac{a}{2 \Delta}\right)^2 \tag{25c}
\end{align}
Equation (25b) uses $(4\pi\mathbf{j}/c) = \nabla \times \mathbf{B}$. 
Dividing Equation (25b) by $({B_{\phi}}^2/8\pi\Delta)$ gives   Equation (25c). 
The far right hand side of Equation (25c) uses flux conservation to write $2 \pi r \Delta B_r = \pi a^2 B_a = \pi a r B_{\phi}$ giving $B_r /B_{\phi}  = (a/2\Delta)$. 

Thus the onset of DCLC changes the dynamics in the following sequence:  DCLC acceleration sets in at $r > R_{ac}$; the Larmor radius $r_L$ increases to equal $\Delta = a/2$, the initial flux width in the nose; and any further increase in energy could cause the ions to escape. 
But, in trying to escape, ions attached to field lines begin to spread out the flux, causing the flux width $\Delta$ to expand so as to contain the most energetic ions. 
And as the flux expands, $\beta_{\perp}\rightarrow 1$ by Equation (25c). 
Thus, in fairly short order, system variables evolve to:
\begin{align}
\Delta &\approx (r_L)_{\text{MAX}} \approx r (E_{\text{MAX}}/eV) \tag{26a} \\  
E_{\text{MAX}} (r) &\approx	E_a + \int_{R_{ac}}^r e \left(E_r\right)_{\text{accel}} \, dr	<	f_{\text{conv}}(eV) \tag{26b} \\
\beta_{\perp} &\approx 1 \tag{26c}
\end{align}
where $E_a$ is the energy as ions emerge from the Central Column if we neglect kink mode acceleration in the nose, and $E_{\text{MAX}}$ is the energy of ions that enter the nose directly from the Central Column, as opposed to cold ions recycling from the ambient as discussed in Section 4.4.

At $\beta_{\perp}\approx 1$, field lines begin to untwist, giving $B_r \ll B_{\phi}$ so that the 
current carried dominantly by $j_r$ can no longer flow parallel to field lines. 
Yet this current must be maintained to satisfy our mean-field MHD equilibrium, since the displacement 
current can be ignored for this slowly evolving field, giving $\nabla \cdot \mathbf{j} = 0$ so that 
the poloidal current $j_z  \rightarrow j_r$ is continuous across the nose. 
Since the nose ion current must flow perpendicular to the dominant field $B_{\phi}$, as field lines untwist, 
a hyper-resistive diffusive transport rate $\langle D^H_{r} \rangle$  is required to carry the current, giving:
\begin{align}
j_r &= - e \langle D^H_{r}\rangle \frac{\p n}{\p r} = en\langle v \rangle \tag{27a} \\
\langle v \rangle 	&=	- \langle D^H_{r} \rangle \left(\frac{1}{n} \frac{\p n}{\p r}\right) \tag{27b} \\
j_z \text{ (Central Column) } &= n e c	\rightarrow	j_r = e n c (\Delta/r)  \quad\text{(nose)} \tag{27c} 
\end{align}
That a net ion current results in Equation (27a) arises from the fact that $\langle D^H_{r}  \rangle$  
by DCLC  turbulence represents a random walk in momentum due to ion scattering analogous to collisional scattering. 
Because DCLC is resonant at the ion cyclotron frequency, scattering by DCLC affects ions only but does not affect electrons. 
That drift waves excited by DCLC do affect both ions and electrons plays a role in vertical transport 
in and out of the nose, in Section 4.4. 
Finally, Equation (27c) takes note of the transition from current density $nec$ in the Central Column 
where current flows along field lines, to a reduction by a factor $(\Delta/r)$ when finally diffusion 
must transport current perpendicular to $\mathbf{B}$, where we anticipate $D^H_{r}  = D^H_{z}  
= \Delta c$ from Equation (29e) below. 

It remains to verify that ion radiation can be neglected in the nose. We rewrite the momentum 
Equation (14a) to apply to the nose, giving:
\begin{align}
\frac{dp}{dt} &=	m_i c^2 \frac{\p \gamma_L}{\p r} = e{(E_r)_{\text{accel}} -  \frac{2}{3} e^2 {\gamma_L}^4 \left[\left(\frac{\left(v_{\perp}/c\right)^4}{{r_L}^2}\right)  +  \left(\frac{1}{r^2}\right)\right]} \tag{28}
\end{align}
where we approximate $\frac{d}{dt} = c \frac{\p}{\p r}$ and the radiation term, from Equation (18), 
represents both synchrotron radiation ($\propto \frac{1}{{r_L}^2}$) and curvature radiation ($\propto \frac{1}{r^2}$). 
Before DCLC sets in, $\left(v_{\perp}/c\right) = \left(r_{L0}/a\right)$ in Equation (23b) 
gives synchrotron radiation equal to curvature radiation. 
Then, even with $\left( E_r \right)_{\text{accel}} = 0$, integrating Equation (28) 
gives $\gamma_L = \gamma_{CC} /\left[1 + 10^{-8}\left(1 - \left(\frac{a}{r}\right)\right)\right]^{1/3} 
\approx \gamma_{CC}$ for any $r < R_{ac}$ before DCLC begins.
When DCLC does set in, the dominant synchrotron radiation with $(v_{\perp}/c) \approx 1$ 
is still negligible compared to DCLC acceleration approximated as $\left( E_r \right)_{\text{accel}} 
\approx V/[r \ln (R/a)]$, giving $E_{\text{SYN}}/E_{\text{accel}} < 0.01$ near $r = R_{ac}$, decreasing as $r$ increases.

\subsection{DCLC Quasi-Linear Transport, Ion Energy Distribution}

We now try to calculate the likely nonlinear outcome of DCLC instability in producing the accelerated ion distribution. 
The non-linear development of DCLC instability was shown to explain momentum transport in the 2XIIB mirror experiment \citep{berk1977}. 
Later \citet{smith1983} applied Kaufman's formal method 
\citep{kaufman1972} 
to derive an exact quasi-linear DCLC transport equation in action-angle space, a method that can be extended to give a Fokker-Planck equation including both diffusion and friction \citep{fowler2007}. 
For relativistic DCLC, the important action variables are the ion magnetic moment $P_{\mu} = \left(p_r c^2/eB_{\phi}\right) = r_{L i} c$, and the radial canonical momentum $P_r = p_r + (e/c) A_r$ with $p_r = m_i \gamma_{\mathrm{L}i} v_r$ and vector potential $A_r(r,z)$ giving $\frac{\p A_r}{\p z} = B_{\phi}$. 
Here we will approximate the formal results in the two-dimensional $P_{\mu}$ and $P_r$ space to proceed directly to a 3D quasi-linear equation for transport in $p_r$, $r$ and $z$, as follows. 

Here and hereafter, we drop the superscript $H$ so that, for example, $D_{p r}$ will mean hyper-resistive diffusion in the radial component of momentum, and so on. We obtain:
\begin{align}
\frac{\p f}{\p t} &= - e E_r \frac{\p f}{\p P_r}  +  \frac{\p}{\p P_r} D_{p r} \frac{\p f}{\p P_r} +  \frac{\p}{\p P_{\mu}} D_{\mu} \frac{\p f}{\p P_{\mu}} + S \tag{29a} \\
	&=  \frac{\p}{\p p_r}\left[- e(E_r)_{\text{accel}} f + D_{p r} \frac{\p f}{\p p_r}\right] + \frac{\p}{\p z} D_z \frac{\p f}{\p z} + \frac{\p}{\p r} D_r \frac{\p f}{\p r} + S\tag{29b} \\
D_{p r} &\approx \langle e E_{1 r} \int_{-\infty}^t	 dt' \, e E_{1 r}(r'(t')) \rangle \tag{29c} \\
D_z 	&=	D_{p r} /\left(\frac{\p P_r}{\p z}\right)^2 = (eB_{\phi}/c)^{-2} D_{p r} \tag{29d} \\
D_r 	&= D_{pr}/(\partial P_r/\partial z)	D_{\mu}
\leq {r_{L i}}^2 \omega_{ci}	\approx	c\Delta	\tag{29e}
\end{align}
Here the source $S$ represents input from the Central Column (at energy $E_a$) plus recycling from the ambient discussed in Section 4.3. 
 In Equation (29a), the first term on the right hand side represents ion acceleration in the 
nose. 
The other terms describe diffusion during acceleration. 
Equation (29b) applies the transform to variables $p_r$, $r$ and $z$. 
For DCLC, the formal theory shows that all transport processes arise from the electrostatic perturbation $E_{1 r}$ giving the relationships shown in Equations (29c), (29d) and (29e). 
Though $D_r$ and $D_z$ turn out to be of similar magnitude, physically $D_r$ derived from $D_{\mu}$ is analogous to a random walk by ion ``collisions'' due to cyclotron resonance, while $D_z$ includes also random walk by DCLC drift-wave $\mathbf{E} \times \mathbf{B}$ motion. 
For strong instability with growth rate $\omega_{c i}$, both processes give a correlation time $\omega_{c i}^{-1}$ with the ion Larmor radius as the step size. 
That $r_{L i}$ is the step size for $\mathbf{E} \times \mathbf{B}$ motion follows from $(cE_r/B_{\phi}\omega_{c i}) \approx r_{L i}$.

We approximate the solution of Equation (29b) by: 
\begin{align}
f(r,E) &\approx n(r)\left[C^{**} \exp \int_{p_a}^{E/c}dp_r[e(E_r)_\mathrm{accel}/D_{pr}]\right]  \tag{30a} \\
&\approx n(r)\left[C^{**} \exp \int_0^EdE'\left(1/E_\mathrm{MAX}(r)\right)\right]  \tag{30b} \\
&\equiv  n(r)	\left(1/1.73E_{\text{MAX}}\right) \exp \left(E/E_{\text{MAX}}(r)\right),\ \ \ E^*<E_a<E<E_\mathrm{MAX}(r) 	\tag{30c}
\end{align}
We will return to the meaning of $E^*$ in Section 4.4.
 
We obtain Equation (30a) by setting the first term in Equation (29b) equal to zero, integrating and factoring out the ion density $n(r)$. We account for $S$ by taking the lower integration limit $p_a = E_a/c$, to account for rapid acceleration of recycling ambient ions up to the energy $E_a$  of Central Column ions entering the nose. And we have set the spatial diffusion terms in Equation (29b) $(\propto (n/e)\mathbf{\triangledown}\cdot \mathbf{j})$ equal to zero. 
In Equation (30b), we first write $dp_r = dE/c$ for relativistic energies and then apply Equations (21b) and (29d), giving $(e/c)(E_r)_{\text{accel}} = (e/c)(D_z/c\Delta)B_{\phi} \approx (D_{p r} /E_{\text{MAX}})$ using also Equation (26a) with $(V/rB_{\phi}) \approx 1$ using $r B_{\phi} = a B_a$ in Equation (9g).
To determine $C^{**}$, we note that DCLC momentum diffusion only spreads energies downward (to fill the ``loss cone'') so that ion energies at radius $r$ do not exceed $E_{\text{MAX}}(r)$, leading us to set $C^{**} = (1/1.73E_{\text{MAX}})$ to give $\int_0^{E_{\text{MAX}}} dE \, (f/n) = 1$. 
This downward spread of energies to achieve a state of marginal stability has been well documented in fusion mirror devices 
\citep{baldwin1977,smith1983}; 
but $(E_r)_{\text{accel}} \propto D_{p r}$ inhibits downward energy spread in jets. 
Note that the final result for $f_0(E)$ only depends on ratios of transport quantities, not their absolute value.



\section{COSMIC RAY ENERGY SPECTRUM, INTENSITY ON EARTH}
\label{sec:CRs}

In this Section, we use the transport equation in Section 3.4 to compare our model with observations of UHE cosmic rays. 

\subsection{Cosmic Ray Energy Spectrum}

In our model, cosmic rays are ejected via the $D_z$ term in Equation (29d). 
That this gives an energy spectrum similar to observations can be seen as follows. 

Since UHE cosmic rays travel great distances, we treat AGN jet/radiolobes as point sources described by integrating the $D_z$ term in Equation (29b) over the entire volume of the nose, giving an energy spectrum $I(E)$. 
As several steps are required, we display them as follows:

\begin{align}
I(E) &= \int_{R(E)}^R 2 \pi r \, dr \int_0^{\Delta} dz \frac{\p}{\p z} D_z \frac{\p f}{\p z} \tag{31a} \\
&\approx \int_{R(E)}^R 2 \pi r \, dr \int_0^{\Delta} dz D_z \left(I/e\langle v \rangle 2 \pi r \Delta\right) f_0(E) \left(\frac{\kappa^*}{\Delta^2}\right) \tag{31b} \\
&\approx \kappa^*(I/e) \int_{R(E)}^R r \, dr \left( \frac{D_z}{r \langle v_r \rangle} \right) f_0(E) \left(eV/r E_{\text{MAX}}\right)^2 \tag{31c} \\
&= \kappa(I/e) \int_{R(E)}^R \frac{1}{r} dr \, f_0(E) \left(eV/E_{\text{MAX}}\right)^2 \tag{31d} \\
f &\approx	n f_0(E) = \left(I/e\langle v \rangle 2 \pi r \Delta\right) f_0(E) \tag{31e} \\
E &= \int_{R(E)}^r dr \, e E_r \tag{31f}
\\
\kappa &= \kappa^*(D_z/D_r) \tag{31g}
\end{align}
with $E_{\text{MAX}}$ given by Equation (26b). 

In Equation (31b), we approximate $\frac{\p}{\p z} = 1/\Delta$ together with a constant $\kappa^*$ that we will use to guarantee energy conservation, as discussed below. 
The lower integration limit $R(E)$ is defined by Equation (31f), chosen to eliminate values of $E$ not accessible by acceleration by $E_r$. 
In Equation (31b), $f$ is given by Equation (31e) taken from Equation (30c), with $n(r) = I/eA<v>$ in Equation (9f) with area $A = 2\pi r\Delta$ for flux width $\Delta$ in the nose, as in Section 3.3.1. Taking $n_i\propto I$ follows since DCLC only drives ion current to maintain current continuity around the loop.
In Equation (31c), we first integrate in $z$ to cancel one factor $\Delta$, then substitute $\Delta = r(E_{\text{MAX}}/eV)$ from Equation (26a) in remaining factors. 
In Equation (31d), we set $(D_z/r\langle v \rangle)= (D_z/D_r)$ using $\langle v \rangle = D_r/r$. 
While ($D_z/D_r$) can be order unity as discussed in Section  3.4, we choose to absorb ($D_z/D_r$) into $\kappa^*$ giving here and hereafter the new fitting parameter $\kappa$ in Equation (31g).

To see that Equation (31d) yields an approximate power law, we note that $f_0$ depends on $(E/E_{\text{MAX}})$ and change variables from $r$ to $Y \equiv (E/E_{\text{MAX}})$, using:
\begin{align}
\frac{dY}{dr} &= -\frac{E}{{E_{\text{MAX}}}^2}\frac{dE_{\text{MAX}}}{dr} \approx -\frac{Y}{r} \tag{32a} \\
dr &= -dY \, (r/Y) \tag{32b} \\
Y(R(E))	&= 1,	\qquad Y(R) = E/eV \tag{32c}~~,
\end{align}
where $V = 1.4\times 10^{20} M_8^{1/2}$ volts is the full dynamo voltage from Equation (9g) with numbers in Section 2.8.
We now apply this transformation to Equation (31d) with $r = R$ as the upper limit of integration and 
$f_0 = \left(1/1.73 E_{\text{MAX}}\right) \exp \left(\frac{E}{E_{\text{MAX}}}\right)$ from Equation (30c).
We obtain:
\begin{align}
I(E) &=	\kappa(I/e) \int_{E/eV}^1 dY \, (r/Y) \frac{1}{r} (1/1.73 E_{\text{MAX}}) (eV/E_{\text{MAX}})^2 \exp Y \tag{33a} \\
&=\kappa(I/e)(1/eV)(eV/E)^3 K(E)
\tag{33b} 
\end{align}
The factor $K(E)$ is given by:
\begin{align}
 K(E) 	&= (1/1.73) \int_{E/eV}^1 dY \, Y^2  \exp Y \notag \\
		&= (2.73/1.73) \left\{1 - \left[2(1 - E/eV) + (E/eV)^2\right] \exp[ - (1 - E/eV)]\right\} \tag{34}
\end{align}
The dominant scaling factor $1/E^3$ arises naturally from diffusion across a flux width $\Delta \propto E_{\text{MAX}}$ and the normalization of $f_0(E) \propto E_{\text{MAX}}^{-1}$ together with the change of variable whereby an integration in $r$ from $R(E)$ becomes an integration in $Y$ from $E/eV$. 
\subsubsection{The Fitting Parameter $\kappa$}
We determine the fitting parameter $\kappa$ by energy conservation. We delay discussion of the physical significance of $\kappa$ to Section 4.4.

Neglecting 
the small power dissipated in the Central Column (1\% by Equation (17)), 
the conservation of the jet power escaping as cosmic ray ions is given by:
\begin{align}
 f_{\text{conv}} IV &\geq \int_{E^*}^{eV} dE \, E I(E) \notag \\
 &=	\int_{E^*}^{eV} dE \, eV (E/eV) [\kappa(I/e)(1/eV)(eV/E)^p] \notag \\
 &=	\kappa IV (p - 2)^{-1}[(eV/E^*)^{p-2}  -  1] \tag{35}
\end{align}
\begin{align}
\kappa &= \left\{\left(p - 2\right)/\left[(eV/E^*)^{p-2}  -  1\right]\right\} f_{\text{conv}} \tag{36}
\end{align}
where we have approximated $(1/E^3)K(E) \approx (1/E^p).$
 
We estimate the conversion efficiency $f_{\text{conv}}$ as follows. 
We approximate $E_r$ by an average value $\langle E_r  \rangle  = [(V/r)/\ln\left(\frac{R}{a}\right)]$ 
which guarantees that $\int_a^R dr \, \langle E_r  \rangle \leq V$ across the nose. 
This same approximation can be used to estimate the efficiency $f_{\text{conv}}$ for 
converting electromagnetic energy into kinetic energy, from Equation (26b) with $(R_{ac}/a)$ in Equation (24c). 
We obtain: 
\begin{align}
f_{\text{conv}} &=	E_{\text{MAX}}(R)/eV =  \ln \left(R_{ac}/a\right)/\ln\left(R/a\right)\approx  0.5 \tag{37a}
\\
E_{\text{MAX}}(r
) &\approx
\int_{R_{ac}}^{r} dr' \, e \left(\frac{V}{r' \ln\left(R/a\right)}\right) =eV[\ln(r/a)/\ln(R/a)] \tag{37b}	
\end{align}

 \subsection{Minimum Energy $E^*$}

As we noted at the end of Section 3, the fact that DCLC is driven by a hole in momentum space produces the peculiar result that DCLC momentum diffusion is not isotropic but mainly tries to fill this hole. At any radius $r$, this
limits the energy distribution to $E^*<E<E_\mathrm{MAX}(r)=e\int_a^r E_r$, where $E^*$ adjusts to maintain marginal stability and $E_\mathrm{MAX}=eV$ ($V$ is the disk voltage) at $r=R$.

Initially $E^*=E_a$, the dominant energy of ions entering the nose from the Central Column, $E_a=m_ic^2\gamma_{CC}=3.4\times 10^{16}M_8^{5/8}$ eV by Equation (19b). DCLC commences where the ion Larmor radius at energy $E_a$ satisfies the DCLC onset condition at marginal stability, given by the equality in Equation (23c). That is, DCLC onset does not determine $E_a$. Rather, $E_a$ determines the radius of onset (giving $r = R_{ac}$ in Equation (24c)). There is no DCLC turbulence at $r < R_{ac}$. On the other hand, as DCLC builds up the plasma pressure parameter $\beta_\perp$, the field at $r = R_{ac}$ is depressed, causing the Larmor radius to increase. In response, DCLC momentum diffusion at $r = R_{ac}$ pushes proton energies downward just enough to re-establish marginal stability.
As DCLC growth causes the pressure parameter $\beta_\perp$ to grow by Equation (25c), the field inside the nose decreases as $(1 − \beta_\perp)^{1/2}$, giving a lower DCLC threshold energy  $E^* = E_a(1 - \beta_\perp)^{1/2}$. We obtain:
\begin{align}\nonumber
 (E^*/E_a)
 &= (1-\beta_\perp)^{1/2} = (a/2\Delta_\mathrm{MAX})= (a/R_{ac})(\mathrm{eV}/E_\mathrm{MAX})
\tag{38}
\end{align}
Here we have applied $\beta_\perp$ in Equation (25c) with $\Delta=\Delta_\mathrm{MAX}$ in Equation (26a) at $r = R_{ac}$; then introducing
magnitudes $(R_{ac}/a) = 1.7 \times 10^5M_8^{-1/5}$ from Equation (24c) and $E_\mathrm{MAX} > E_a$ . More accurate determinations of DCLC marginal stability and the role of an additional Alfven Ion Cyclotron (AIC) instability occurring at high $\beta_\perp$ are discussed in \citet{fowler1981} using non-relativistic theory, giving with relativistic masses the same results as relativistic DCLC theory, as discussed in Appendix B2.

\subsection{Calibrating the 
 Model $I(E)$ to Observations}



We compare our $I(E)$ with Figure 1 in \citet{cronin1999}. 
This curve translates the count rate on Earth to an intensity proportional to our $I(E)$. As Cronin notes, like our $I(E)$, the basic scaling is $(1/E^3)$, then reducing to $(1/E^{2.7})$ at higher energies, perhaps for other reasons mentioned by Cronin. Cronin conveniently relates the plotted intensity to count rate above a specified energy giving $I_1\propto \int_{E_1}dE\ I(E)\propto(1/E_1^2)$ for $I(E)\propto(1/E^3)$. This gave $(1/$km$^2$ year) for cosmic rays above about $5 \times 10^{18}$ eV. The same scaling correctly predicted results giving $(1/$km$^2$ century) above $6 \times 10^{19}$ eV for extra-galactic sources detected at the Pierre Auger Observatory \citep{pierre2007,pierre2014,pierre2017}. Applying Cronin's scaling to our model gives for a single AGN source: 

\begin{align}
I_1(E_1) &=  \int_{E_1}^{eV}dE \, I(E) =  \int_{E_1}^{eV} dE \, \left[\kappa (I/e^2 V) (eV/E)^{p}\right] \notag \\
&=(I/e)\left(\kappa/(p-1)\right)\left[\left(eV/E_1\right)^{p-1} - 1\right] 
\tag{39a
}
\\
V &= 1.4 \times 10^{20} {M_8}^{1/2}\text{ volts}\quad;\qquad I = 0.7 \times 10^{28} {M_8}^{1/2}\text{ esu/s} \tag{39b
}
\end{align}
where $I_1$ is the  
ion flow rate 
for all energies $>E_1$, and Equation (38b
) gives $I$ and $V$ from Equations (8b), (8c) and (9d). That $I_1 > I/e$ implies both electron and ion recycling to  maintain charge neutrality and a net disk current $I$ as required by $\triangledown \cdot \mathbf{j}= 0$. Recycling with the ambient is discussed in Section 4.4.	

In terms of $I_1$, data points on Cronin's curve for AGNs at a distance $R_S$ correspond to a flux on Earth equal to $(I_1/4\pi R_S^2).$ To compare with the Pierre Auger result above, we take $E_1 = E_\mathrm{PA} = 6 \times 10^{19}$ eV (\citet{pierre2007}). Summing over $N$ sources gives:



\begin{align}\nonumber
\Sigma_{n=1}^N \left(I_1(E_1)/4\pi R_S^2\right)_n
= 1/(\mathrm{km}^2 \mathrm{\ century})\tag{40a
}
\\ \nonumber
0.1    <  (E^*/E_a)M_8^{13/8}    <    1.0
\hspace{1.5cm}\tag{40b
}
\\ \nonumber
(E^*/E_a)= (1-\beta_\perp)^{1/2}\hspace{1.5cm} \tag{40c
}
\end{align}


The range in $E^*$ in Equation (31b) corresponds mainly to the range in $R_S$, from the closest to the most distant AGNs accessible at the Pierre Auger Observatory (470 AGNs within a 
240 Mlyr 
range, \textcolor{blue}{Pierre Auger Collaboration 2007}). As examples, the lower limit in Equation (31b) would correspond to a single nearby source ($N = 1$, $R_S$ = 10 MPc) and the upper limit corresponds to several distant sources ($N = 10$, $R_S$ = 100 MPc).

How $E^*$ in our model can be consistent with Equation (40b
) concerns density profile adjustments to maintain marginal stability to DCLC. We employed marginal stability to determine the onset of DCLC by taking the equality in Equation (23c
) for ions of energy $E_a$ entering the nose from the Central Column. The onset of DCLC causes the plasma pressure parameter $\beta_\perp$ to grow.

As $\beta_\perp$ grows, the field inside the nose decreases as ($1-\beta_\perp)^{1/2}$, giving a lower DCLC threshold energy $E^∗$ given by Equation (40c). Equation (40c) could accommodate Equation (31b), depending on the uncertain value of $\beta_\perp$ near the DCLC onset at $r = R_{ac}$. More accurate determinations of DCLC marginal stability and the role of an additional Alfven Ion Cyclotron (AIC) instability occurring at high $\beta_\perp$ are discussed in \citet{fowler1981} using non-relativistic theory, giving with relativistic masses the same results as relativistic theory, as in the case of DCLC (see Appendix B.2).

\subsection{Recycling with the Ambient
}
The small value of $\kappa$ in Equation (36
) concerns recycling of protons as the jet nose pushes against the ambient. Recycling is given by:
\begin{align}
I_{\text{RECYCLE}} =  e \int_{E^*}^{eV} dE \, I(E) - (I/e) \approx (I/e) \left(\frac{\kappa}{p-1}\right) 
\left[(eV/E^*)^{p - 1} - 1\right]  \approx (I/e)\left[\frac{1}{2}f_\mathrm{conv}(\mathrm{eV}/E^*)\right]
\tag{41a}
\\
P_\mathrm{CR}(r)	=	\int_{E^*}^{E_\mathrm{MAX}} dE\ E I(E)	=      IV (f_\mathrm{conv} E_\mathrm{MAX}(r)/eV) \hspace{2.5cm}\tag{41b}
\end{align}

In Equation (41a), we use $\kappa$ from Equation (36), giving  a result independent of the power law exponent $p$.  Equation (41b) gives the distribution of cosmic ray power $P_\mathrm{CR}$ across the radial profile of the jet nose, with lowest energy ions nearest to the tip of the jet. 

We note that the necessity for recycling is characteristic of a decreasing power law $(C/E^p)$. An excess of ions beyond that needed to conduct the jet current $I$ results for any specified total power $IV$ applied to the conservation of energy (to calibrate $C$),
then 
applied 
to the conservation of particles (giving $I_\mathrm{REC} \approx (I/e)\{ [(p-2)/(p-1)][(V/E^*) - 1]\}$. Our model makes this quantitative by showing that the $(1/E^3)$ power law is characteristic of a single AGN source whose black hole mass $M$, together with angular momentum conservation, determines all jet parameters as shown in Section 2.8.  Then the total emission of extra-galactic cosmic rays includes some reaching Earth, some not.
	
We note also that the ambient can sustain recycling and recycling does not appear to change our model parameters.  
%
Physically, recycling is part of DCLC vertical transport, whereby adjustments in the Vlasov distribution functions $\partial f(x,p)/\partial z$ for hot ions, ambient ions and electrons do in detail what adjusting $\kappa$ does in our model. Thus recycling commences where DCLC commences, at $r = R_{ac}$ in Equation (24c).  
%
%
Because of the low jet density in Equation (9f), snow-plowing an ambient density $n_\mathrm{amb} \approx 0.01 n_\mathrm{jet}$ over an annular width $\delta r < R_{ac}
$ at $R_{ac}$ would be sufficient to accommodate the result in Equation (41a).
The jet density $n$ in Equation (9f) already includes recycling, any excess being expelled vertically by DCLC diffusion $D_z$ to maintain just the ion density needed to conduct the current sustained by radial DCLC diffusion $D_r$. Nor would recycling 
add much to ambient pressure
, the ratio being $[(n E)_\mathrm{RECYCLE}/(n_\mathrm{amb} m_i (dL/dt)^2] < 1
$ using $n_\mathrm{RECYCLE} \approx (10^5 I/ec\pi R^2) \approx 100 n_\mathrm{amb} (a/R)^2 \approx 10^{-15} n_\mathrm{amb}$ and $m_i (dL/dt)^2 =  10^{5}$ eV, with $(a/R)^2 \approx 10^{-17}$.

\section{
ELECTRON RADIATION, PAIRS, FLARES, NEUTRINOS
}
\label{sec:ele}
This Section is mainly concerned with what we can say concretely about electron  
radiation in  quasi-steady state. We will focus on electron synchrotron radiation extending to microwaves, which for decades has been and continues to be the main evidence for the jet/radiolobe structures that we claim also create the UHE cosmic rays reaching Earth. 

We will neglect weak DCLC acceleration of electrons in the nose and focus on the Central Column, where electrons are accelerated to TeV's, by Equation (19b). With steady state in mind, we 
omit discussion of transient flares
, perhaps byproducts of kink mode turbulence. We neglect dilution of ions by positrons which, by two-stream interaction with their electron partners, cannot contribute to the jet current. Finally, we omit discussion of neutrinos, but note that the essentially mono-energetic proton beam $(\approx 3 \times 10^4\mathrm{\ TeV})$ produced by kink mode acceleration in the Central Column has associated
with it lower energy protons following wandering field lines. As is mentioned in Section 5.4, these wandering protons accelerated mainly when they pass through the Central Column can spread proton energies downward, from $3 \times 10^{16} M_8^{5/8}$ eV  to TeVs, while protons accelerated near $r << a$ where field line curvature is small would add a tail up to $10^{18}$ eV by Equation (17). 


\subsection{Electron Synchrotron  
Model}

The two main features of our synchrotron model are: (1) angular momentum conservation in 
Equation (6c) which serves to project the magnetic field pattern at the disk all along the jet so that 
Central Column parameters at the disk determine synchrotron wavelengths; and (2) plasma 
turbulence which spreads this field projection over an ever-widening cone like the dashed 
path in Figure \ref{Fig1}, finally expanding into giant 
radiolobes like the closed flux in that figure. 

As noted earlier, what causes some field lines to wander away from the 
Central Column is MHD kink instability. 
As is explained in detail in Paper II, there are two distinct kinds of MHD kink modes driven 
by the jet current, confirmed in MHD simulations mentioned in Section \ref{sec:evidence}. 
In Paper II, we derived the two kinds of kink modes from the MHD energy principle 
giving the free energy for magnetic perturbations of the form 
$\delta W \propto \int r \, dr \, \bm{\xi} \cdot \left(\mathbf{j}_1 \times \mathbf{B} 
+ \mathbf{j} \times \mathbf{B}_1\right)$ where subscript 1 denotes 
perturbations and $\bm{\xi} \propto \exp i\left[m \phi + k r\right]$ is a displacement of magnetic field lines. 

Internal kink modes with $k a \approx 1$ localized to the Central Column 
create the well-understood kink mode accelerator in Section 3. 
Internal kink modes are characterized by magnetic resonances in $\delta W$. 
External kink modes of interest here are those with $k$ small enough to 
avoid these resonances, requiring $k R_0 \leq 1$ for $R_0$ shown in Figure \ref{Fig1}, 
also dominated by $m = 1$ describing a rigid body motion of the entire magnetic structure. 
The other point made in Paper II is that, because 
accretion disk power is concentrated near $r=a$, $\mathbf{\xi}$ should
be thought of as arising at $z\approx0$, $r\approx a$, growing not in time but
in $z$ along the jet length; and relativistic effects give ``secular,'' 
not exponential, growth, giving $\xi \propto z$.

Thus we come to a picture of the synchrotron radiator localized in the Central Column 
together with an array of magnetic field lines emerging from the disk as the dotted cone in 
Figure \ref{Fig1}, ``wandering'' in the sense of a twist of the 2D structure as a whole, 
first near the black hole but growing radially along the jet when viewed in 3D. 

We will be able to predict the synchrotron power and dominant wavelengths without detailed 
knowledge of the wandering field line structure. The power is known from the calculation of the 
voltage drop in Equation (17). The magnetic field strength determining synchrotron wavelengths 
can be determined by angular momentum conservation, as follows.

Angular momentum projects the accretion disk field forward, giving synchrotron radiation at 
approximately the same magnetic field strength all along the conical field line structure.
This can be seen by the following extension of Equation (6c) into the jet: 
\begin{align}
1/2\dot{M} \Omega_a = a {B_a}^2 & = \int_0^z dz \langle r|\mathbf{B}_{\text{POL}}|\cdot\vec{\triangledown}B_\phi \rangle_\phi \approx \langle r\delta B_\mathrm{POL}\delta B_\phi\rangle_\phi  \tag{42a} \\  
|\delta B_\phi | & = k\xi B_\phi(r) = (k\xi)(a/r)B_a  \tag{42b} \\
\delta B_\mathrm{POL} & \approx (k\xi)^{-1}B_a \approx B_a \tag{42c}~~. 
\end{align}
Here $\frac{1}{2}M^* \Omega = a {B_a}^2 $ by Equations (8d) and (9d). We extend the meaning of $dz$ in Equation (42a) to include paths along wandering field lines, averaged over $\phi$ as indicated by $<...>_\phi$ giving correlated perturbations on the right hand side, omitting small mean field 
 contribution at $r > a$, giving the Diffuse Pinch in Figure \ref{Fig1}. Equation (42b) 
 is the MHD perturbation for wandering field lines at $r > a$ derived from Faraday's Law, as  in deriving  $\delta W$ above. Applying Equation (42b) to Equation (42a) gives Equation (42c). 
We set $k\xi\approx 1$ as is typically observed in simulations, 
still obeying $|\delta B_\phi|<B_\phi(r)$ at all r.
 
The path integral giving Equation (42a) is also the path of synchrotron-emitting electrons following 
field lines that leave the Central Column. 
We conclude that, although the mean field $B_o$ falls off as $1/r$ outside $r > a$, the total field $B = B_o + \delta B_\mathrm{POL}\approx B_a$ wherever accelerated electrons flow within the dashed cone in Figure \ref{Fig1}, yielding 
the wavelength distribution in Section 5.3. 
Poloidal flux is properly conserved by Fourier  expansion, giving the mean field $B_0 = <\sum_m B_m \exp (im\phi)>_\phi$ that is uniquely the Grad-Shafranov solution in Figure \ref{Fig1} containing all of the flux (see Paper II). Also $B_1$ of order $B_0$ was confirmed in the tower-like MHD simulations in 
\citet{carey2011},
giving $B_1 = 0.4B_0$ in quasi-steady state. We have ignored the build up of flux-amplifying closed structures, as in certain fusion devices, negligible for astrophysical jets in which helicity injection by the accretion disk is mainly consumed in lengthening the jet \citep{fowler2009}. 

\subsection{Synchrotron Luminosity}

The synchrotron radiation described in Section 5.1 dissipates a large part of the electron energy produced by kink mode acceleration in the Central Column. 
Electron acceleration is equal and opposite to ion acceleration in Section 3.2, giving $\frac{1}{2} I \Delta V \approx 0.005 IV$ as the synchrotron power, or luminosity, in good agreement with observations \citep{krolik1999}. 
The synchrotron power is given by $I$ and $V$ in Equation (8), yielding:
\begin{equation}
P_{\text{SYN}} \le\frac{1}{2} I \Delta V \approx 0.005IV \approx 1.5 \times 10^{43} M_8\text{ erg/s} \tag{43}~~.
\end{equation}
where the inequality accounts for non-synchrotron electron radiation \citep{li2000}. This luminosity is consistent with observations of powerful AGN jets \citep{krolik1999}.  


\subsection{Synchrotron Wavelengths}

As noted above, a new feature of our model is that, both in the Central Column and in the surrounding 
structure (the cone and radiolobe regions in Figure 1), synchrotron radiation should always be calculated 
using the magnetic field $B_a = 1500 M_8^{-1/2}$ Gauss at $r = a$ near the black hole. 
Then, wherever they occur, the frequency $\nu$ and wavelength $\lambda$ for synchrotron emissions at the n$^\mathrm{th}$ harmonic are given by:
\begin{align}
\nu= n(eB_a/2\pi m_e c\gamma_\mathrm{Le}) = 4\times 10^9(n/\gamma_\mathrm{Le})M_8^{-1/2} \text{ Hz}\tag{44a} \\
\lambda= (c/\nu) = 7 (\gamma_\mathrm{Le}/n) M_8^{1/2} \text{ cm} \hspace{1cm}\tag{44b}
\end{align}
with numbers from Section 2.8. 

For a given electron energy $m_ec^2\gamma_\mathrm{Le}$, radiation intensity is spread over a range of frequencies around the fundamental $n=1$ (\textcolor{blue}{Jackson\ 1998}, Equation 14.31), the total radiation for all harmonics being that given by Equation (18) applied to electron cyclotron orbits in a magnetic field. Observed radiation can be found by integrating the radiation at a fixed $\gamma_\mathrm{Le}$. An example of this procedure for an assumed energy distribution is given in \citet{li2000}.

In  principle, the electron transport model in Appendix C.3 could yield a correct distribution, but this has proved more difficult than the corresponding ion calculation in Section 3.4; and our techniques do not yet treat flares and hot spots except to note that transients consume less power than steady state \citep{fowler2009}. 
That this does not matter for our main goal of explaining UHE cosmic rays follows from the limited power in all electron radiation, by Equation (43). 

We do predict an upper limit on electron energies $>$ TeVs due to curvature radiation in the Central Column, giving in Equation (19b) $\gamma_\mathrm{L}=\gamma_\mathrm{CC}=3.4\times 10^7M_8^{5/8}$ both for electrons and ions in the Central Column. Our estimate of lower ion energies outside the Central Column, in Section 5 .4, suggests the same for electrons, indicating the same maximum synchrotron wavelengths everywhere in the jet/radiolobe system (the upper limit in Equation (44b)) but the shortest wavelengths and highest frequencies in and near the bright Central Column with radius $a=10R_a$ by Equation (9a).  


\subsection{Synchrotron Opening Angle}

As is mentioned in Section \ref{sec:evidence}, wandering field lines due to MHD kink modes distribute 
synchrotron radiation power over the dashed cone in Figure \ref{Fig1}, though the axi-symmetric 
Central Column remains magnetically collimated. In Paper II, it is shown that relativistic ion 
acceleration on kinking 
field lines yields secular growth of the field line displacement $\xi$, giving then an opening angle 
of the dashed cone in Figure \ref{Fig1} given by:
\begin{align}
\Theta \approx \frac{\xi(z)}{z} = \left(c/z \gamma_0\right)t = 100/\gamma_0\approx 10(R_0/L) \approx 0.01 \tag{45} ~~, 
\end{align}
giving $\Theta$ in radians. 
Here we combine Equations (39) and (53) in Paper II, giving $\gamma_0 \approx 10(L/R_0) \approx 10^4$ as the average Lorentz factor of ions wandering in and out of the kink accelerator zone concentrated in the Central Column. 
An ion with $\gamma_0 \approx 10^4$ for the most extreme field line excursions is to be 
compared with a maximum of order $10^7$ in Equation (19b). 

\section{SUMMARY, COMPARISON TO PREVIOUS WORK}	
\label{sec:sum}

This is the final paper in a series of four papers showing that the production of ultra high energy cosmic 
rays  occurs naturally when processes exist to create an efficient dynamo by accretion around 
a supermassive black hole inside an AGN.
Section 2 and Appendix A argue that MRI is the likely process for such a dynamo.

Given the ubiquitous presence of seed magnetic  fields sufficient to excite a persisting MRI dynamo, 
as discussed in Appendix A.3, our model predicts that all AGNs should produce UHE cosmic rays, 
typically for a jet lifetime $\tau=10^8$ yrs that we equate to the time for accretion to exhaust ambient 
mass around a black hole (accretion rate $M/\tau$). Why actually only $10\%$ of AGNs are the 
radioloud variety we associate with UHE cosmic rays may concern the $\tau=10^8$ year jet 
duration in relation to the 100-times-longer life of the Universe. The fact that $\tau$ is 
independent of black hole mass may imply a connection to the Eddington limit on 
luminosity \citep{beskin2010}. Why some AGNs eject one-sided jets (up or down) should, 
by our model, require a one-sided dissipation of magnetic helicity (a product of poloidal and toroidal fluxes) 
since helicity conservation requires two jets as discussed in Appendix B of Paper I. 

In Sections 3--5, we have shown that the magnetically-collimated jet predicted in Section 2 
automatically becomes a particle accelerator that explains many features of UHE cosmic rays 
and the synchrotron radiation by which AGN jets are observed. A common theme is the use of 
hyper-resistive diffusion in a two-dimensional, mean-field Ohm's Law to evaluate the 
consequences of 3D turbulence, both MHD and kinetic, often avoiding the need to 
know turbulence magnitudes through Onsager-like relationships among observables 
(for example, in deriving the cosmic ray energy spectrum in Equation (33b)). Kinetic effects play key roles: 
to sustain a coronal short-circuit current for the duration of the initial conical jet (Equation (4a)); the two-stream instability, which allows a relativistic kink-mode accelerator to exist in the Central Column (Equation (13c)); and the DCLC instability that drives cosmic ray acceleration in the jet nose (Section 3.3.3).

Our approach differs from conventional wisdom in that, given the dominance of magnetic 
fields above disks if MRI is the creator of a dynamo, we regard accretion disks, jets and the jet nose 
highlighted here as a single magnetic system evolving according to the hyper-resistive Ohm's Law 
given in Equation (2a). 
In this model, the bright jets ejected by AGNs serve mainly as a magnetically-guided conduit 
of power from the black hole dynamo to the nose end of the jet. 
That the jet is magnetically-collimated serves both to stabilize this linkage and also to limit 
energy loss by the synchrotron radiation by which jet/radolobe structures are observed, 
thus leaving most of the jet power to accelerate and eject cosmic ray ions in the nose.

We recognize that magnetic collimation of jets over Mpc dimensions is controversial. 
Three main schools of thought have contributed to understanding magnetized accretion disks and 
the jets they produce. 
One school, stimulated by the centrifugal ejection model of 
\citet{blandford1982}, has focused on magnetic jets as accelerated 
``winds'' \citep{beskin2010}, either at the disk or by conversion of magnetic energy 
to kinetic energy early in the jet history. 
Another school has focused on simulations from first principles 
(GRMHD codes, \citet{mckinney2012}). 
A third school, including our work, has focused on the analogy between 
AGN jets and magnetically-collimated jets produced in the laboratory 
\citep{colgate2004,lovelace1976,zhai2014}. 
 
By using simple models to identify critical physics, we believe that, as summarized in 
Section \ref{sec:towerstructure}, we have closed the loop among these different schools of thought, 
with the conclusion that well-understood laboratory jets are the correct model for AGN jets on 
astrophysical timescales, but only after an initial conical jet launched by rotation has evolved to 
the magnetically-collimated jets required for our model of cosmic ray acceleration. 
Interestingly, the same rotation that makes accretion 
jets different from laboratory experiments at launch eventually causes a jet to 
become magnetically-collimated, already occurring at 1$\%$ of  its final length, 
as estimated in Section \ref{sec:evidence}.  
Other differences from previous work are: 
\begin{enumerate}[(1)]
\item In Paper I: The concept of jets generated by whirling magnetic flux frozen into the accretion disk \citep{frank2002,lynden-bell1996} is replaced by non-ideal MHD in which MRI-driven hyper-resistivity is the mechanism of jet ejection. 
As in magnetized jets created in the laboratory, these jets are ultimately ejected vertically by an electrostatic sheath, with just enough ion mass to carry the current, as opposed to continuing centrifugal ejection 
\citep{blandford1982}. The electrostatic sheath is discussed in Paper II, Appendix A.2.

\item In Paper II: We show that magnetic collimation produces MHD kink instability at short-wavelengths known 
to accelerate electrons and ions in the laboratory without destroying collimation, together with very 
long wavelength kinks producing wandering field lines carrying the power observed as synchrotron radiation 
by electrons in giant radiolobes and as a bright cone with opening angle 0.01 radians, consistent with our model. 

\item In Paper III and this paper: We show that, in the nose-end of the jet, flaring field lines both suppress electromagnetic radiation and create the conditions for kinetic instability that accelerates ions to energies equal to the dynamo voltage. 
Unlike transient mechanisms along the lines of Fermi's initial explanation of cosmic rays, we find that AGN jets become steady-state accelerators that can convert about half of the jet power into the kinetic energy of UHECRs. 
Our acceleration mechanisms are based on known plasma processes that can be further explored in the laboratory as a guide to astrophysical simulations and observations (see Paper III).

\end{enumerate}

Our model predictions are consistent with nine observables, outlined in Table 1 with references to the text in this paper. 
Additional ways to test the model in future work are:

\begin{enumerate}[(1)] 
\item 	In GRMHD and Relativistic MHD simulations of accretion disk jets, apply a diagnostic similar to Figure 
\ref{Fig5} in order to measure directly the short-circuit current predicted in Sections 2.2 and 2.6 as the mechanism of transition from a conical jet to a magnetic tower. 

\item New laboratory experiments on ion acceleration outlined in Paper III.

\item Non-linear kinetic simulations of two-stream instability in a relativistic current-carrying column to verify the persistence of a kink-mode accelerator in the relativistic regime.
\end{enumerate}

\section*{ACKNOWLEDGMENTS}

We wish to express our deep appreciation for the intellectual foresight by our colleague Stirling Colgate that motivated this paper and Papers I and II coauthored with him, still in progress when Stirling passed away December 1, 2013. 
We thank Alexander Tchekhovskoy for suggesting the short-circuit current and Paul Bellan for its theoretical formulation in Equation (2b). We thank Joseph McClenaghan for the use of Figure 6. We also thank Mitch Negus for his help in preparing the manuscript. HL gratefully acknowledges the support of the U. S. Department of Energy Office of Science, the LANL/LDRD program and NASA/ATP program for this work. RA acknowledges the California Alliance Postdoctoral Fellowship Program, the Simons Foundation and the John Templeton Foundation.  

\newpage
 
\section*{Table 1. Model Predictions}
 
\hspace{-0.5cm}\underline{Jet Properties}
\begin{enumerate}[1.]
\item Jet length $L = 10^{24}\text{ cm} = 300\text{ kpc}$		\hfill Equation (10c)
\item Radiolobe radius $R \approx 0.1 L$	\hfill Equation (10d)

\hspace{-0.5cm}\underline{Synchrotron radiation}
\item Synchrotron luminosity \hfill	Equation (43)
\item Synchrotron wavelength $\lambda_{\text{SYN}}=7(\gamma_\mathrm{Le}/n)$ cm \hfill	Equation (44b)
\item Synchrotron opening angle $\Theta \approx 0.01 \text{ radians}$ \hfill	Equation (45)
\item Maximum Electron Energy  $17 M_8^{5/8}$ TeV		\hfill	Equation (19b)

\hspace{-0.5cm}\underline{UHR Cosmic Rays}
\item Maximum UHECR energy $1.4 \times 10^{20} {M_8}^{1/2}\text{ eV}$ 	\hfill Equation (38b
)
\item UHECR intensity on Earth 						\hfill Equations (39a,b
)
\item UHECR energy spectrum  $I_{CR}(E) \propto E^{-p}~~ (p \approx 3)$	\hfill Equation (33b)
\end{enumerate} 

\newpage

\section*{APPENDIX A: MRI-DRIVEN ACCRETION DISK DYNAMOS}

The evolution of conical jets to magnetic towers in Section 2 depends on MHD 
hyper-resistivity $\mathbf{D}$ inside the disk, given by $\mathbf{D} = 
-\frac{1}{c}\langle \mathbf{v}_1 \times \mathbf{B}_1 \rangle$  \citep{fowler2007}. 
A finite mean-field $\mathbf{D}$ emerges if correlations between $\mathbf{v}_1$ and $\mathbf{B}_1$ give a finite value when $(\mathbf{v}_1 \times \mathbf{B}_1)$ is averaged over fluctuations in the same way that gives the mean fields in Section 2, indicated here by $\langle ... \rangle$. 
That conical jets evolve to the magnetic towers of our model follows from Equation (3d) in the main text, requiring $(-D_{\phi} /D_r) = 1$ together with vertical accretion velocity $v_z = (H/r)v_r < v_r$ and $B_r \approx 0$ justified by requiring a field line tilt $\delta r/r \equiv (HB /rB) < 2/3$, at the threshold of the vertical force balance of gravity against centrifugal ejection of mass (Paper I, Appendix A; \citet{blandford1982}).

The proof that $(-D_{\phi} /D_r) = 1$ is derived from:
\begin{align}
(-D_{\phi} /D_r)   &=	|\langle v_{1 z} B_{1 r} - v_{1 r} B_{1 z}  \rangle /\langle v_{1 z} B_{1 \phi}  -  v_{1 \phi} B_{1 z}  \rangle | \tag{A1a} \\
		    &\approx	1 \quad;\qquad v_{1 \phi} \approx  v_{1 r} \quad;\quad B_{1 \phi}   \approx   B_{1 r} \tag{A1b}
\end{align}

In Paper I, Appendix A, $(-D_{\phi} /D_r) \approx 1$ follows from assuming equipartition of $r$ and $\phi$ perturbations in the non-linear steady state $(v_{1 \phi}  \approx v_{1 r}, B_{1 \phi} \approx B_{1 r})$. 
Then, given proper correlations, the numerator and denominator are equal without knowledge of $v_{1 z}$ and $B_{1 z}$. 
As was stated in Paper I, with these assumptions Equation (A1b) is valid for the non-linear steady state for any kind of MHD-like fluctuations.

\subsection*{A.1: Hyper-Resistivity by MRI}

We now evaluate Equation (A1a) using $\mathbf{v}_1$ and $\mathbf{B}_1$ from the linearized equations giving MRI instability in \citet{balbus1998}---hereafter BH98. 
In applying linearized theory, we assume that MRI produces the non-linear correlations necessary to yield a finite $\mathbf{D}$, as kink modes are known to do as discussed in Section 3.2. Then linear theory provides an estimate of ratios of mode amplitudes. 
We obtain: 
\begin{align}
(-D_{\phi} /D_r) &= \left|\frac{\langle \xi_r\left(iv_{1 z} k_z B_z +  i\omega B_{1 z}\right) \rangle}{\langle \xi_{\phi}\left(-iv_{1 z} k_z B_z + i\omega B_{1 z}\right) \rangle}\right| \tag{A2a} \\
 	    &\approx	\left|\frac{\xi_r}{\xi_{\phi}}\right|	=   \frac{\gamma^2 + {k_z}^2 {v_{Az}}^2}{2 \gamma \Omega} \approx \frac{\gamma}{\Omega} \tag{A2b}\\
- i\omega \mathbf{B}_1 &= - c \, i\mathbf{k} \times \left(- \frac{1}{c} \mathbf{v}_1 \times \mathbf{B}\right) = - i\omega (ik_z \bm{\xi})B_z \tag{A2c}\\
\bm{\xi}(r,z,t) &= \bm{\xi} \exp \left\{i \left(k_r r  +  k_z z - \omega t\right)\right\} ; \quad B_{1 r} = i k \xi_r B_z ; \quad B_{1 \phi} = i k \xi_{\phi} B_z \tag{A2d}
\end{align}

Equation (A2b) is the MRI replacement for Equation (A1b), obtained as follows. We apply linearized MRI theory for the simplest case of plane waves $\propto \exp \{ i k_z z - \omega t \}$ in a uniform $B_z$ field, 
with growth rate $\gamma = Im (\omega) \approx k_z v_{Az}$ (BH98, Equation (111)). 
As is common for ideal MHD, we write perturbations in terms of a vector field line displacement $\bm{\xi}$ giving $\mathbf{v}_1 = - i \omega \bm{\xi}$. 
Using this and $\mathbf{B}_1$ obtained from the induction equation, Equation (A2c), we obtain Equation (A2d). Using this and Equation (A4d), we obtain Equations (A2a) and (A2b).

The approximation $(-D_{\phi}/D_r) \approx \gamma/\Omega \approx 1$ applies for the maximum MRI growth rate, $\gamma \approx k_z v_{A z} \approx \Omega$, representing the linearized modes most likely to grow out of the noise. For these modes, with $\gamma = \frac{3}{4} \Omega$ and $k_z v_{Az} = (\sqrt{15}/4)\Omega$ (BH98, Equation (114)), we obtain $(-D_{\phi} /D_r) = 1$ exactly. 
We then assume that fluctuations giving $(-D_{\phi}/D_r) \approx 1$ during peak growth are also characteristic of the steady state \citep{kadomtsev1965}.

\subsection*{A.2: Jet Evolution by MRI}

Consider an initial condition --- typical of GRMHD simulations --- with a pre-existing black hole mass $M$, 
a pre-existing finite poloidal magnetic field $B_z$, but $B_{\phi} = 0$ and an 
ambient environment that we take to be a hydrogen plasma at $1\text{ keV}$ for an actual accretion disk. 
Then the dynamics inside the accretion disk is described by mass flow and the radial force equation and angular momentum equation, giving:
\begin{align}
\frac{\p \rho}{\p t} + \frac{1}{r}\frac{\p }{\p r} r \rho v_r &= 0 \tag{A3a} \\ 
\frac{\p v_r}{\p t} + \frac{MG}{r^2} - r\Omega^2 + \frac{1}{2} \frac{\p {v_r}^2}{\p r} &= \frac{1}{8\pi\rho}( - \frac{\p }{\p r} {B_z}^2 ) \tag{A3b} \\
\frac{\p \Omega}{\p t} +  \frac{v_r}{r^2} \frac{\p }{\p r} \left(r^2 \Omega\right) &= \frac{1}{4\pi \rho r^2} \frac{\p}{\p z} r B_{\phi}B_z \tag{A3c}
\end{align}
where we have used Equation (A3a) to eliminate terms in $\frac{\p \rho \mathbf{v}}{\p t} + \nabla \cdot \rho \mathbf{vv}$, giving then Equation (A3b) on dividing by $\rho$ and Equation (A3c) on dividing by $\rho r^2$. 
We have omitted the viscosity in Eq. (A3c), which
is small to order $(v_r/v_\phi)$ \citep{colgate2014}.
Gravitational accretion begins for the few ions heading directly toward the black hole (the $\frac{\p {v_r}^2}{\p r}$ term in Equation (A3b)) and more importantly, for the much larger population near enough to the black hole so that gravity exceeds their centrifugal force $\propto r \Omega^2$. 

In the limit that $B_{\phi} = |\mathbf{D}| = 0$ in Ohm's Law in Equation (3a), 
accretion is advective, whereby accreting ions stuck to field lines compress the $B_z$ flux, 
possibly creating magnetic stagnation of accretion by the term on the right hand side of Equation (A3b) (true in some GRMHD simulations). 
Then, as $\mathbf{D}$ arises by MRI, the $B_{\phi}$ terms dominate, giving a mainly Keplerian 
disk ejecting a conical jet that transitions to a magnetic tower as discussed in Section 2, 
where $\dot{M} = M/\tau$ and Equation (6c) follow from integrating Equations (A3a) and (A3c) 
in $z$ over the disk half-height $H$ \citep{colgate2014}. 
The timescale for MRI evolution is typically $(10/\gamma) \geq (10/\Omega(r))$. 
The coronal ratio $|B_z/B_{\phi}| = (a/r)^{1/2}$ characterizing magnetic towers is achieved 
as positive $D_r$ makes $B_{\phi}$ grow by Equation (3a) while negative $D_{\phi}$ halts 
advective compression of $B_z$ by Equation (3b).

\subsection*{A.3: Persistence of MRI Dynamos}

For MRI as the dynamo process, the main issue is the persistence of $(-D_{\phi} /D_r) \approx 1$ 
as instability saturates. 
Given an adequate seed magnetic field, MRI can persist even as $B_{\phi}$ grows from 
zero to exceed $B_z$ 
(BH98, Section IV.F). 
To see this, let us revise the dispersion relation in BH98 (Section IV.B), by adding 
$B_{\phi}$ to the equilibrium and $k_{\phi}$ to the perturbation, giving $\mathbf{v}_1 = - i \omega \bm{\xi} \propto \exp \left\{ i \left(m \phi + k_z z - \omega t\right) \right\}$ with $k_{\phi} = m/r$. 
We obtain:
\begin{align}
\mathbf{B}_1 &= i \mathbf{k} \times \left(\bm{\xi} \times \mathbf{B}\right) \tag{A4a} \\
-\omega^2 \xi_r + 2 i\omega \Omega \xi_{\phi} &= -\left[r \frac{\p \Omega^2}{\p r} + \left(\mathbf{k} \cdot \mathbf{v}_A\right)^2\right] \xi_r \tag{A4b} \\
-\omega^2 \xi_{\phi} - 2 i\omega \Omega \xi_r &=	- \left(k_z v_{Az}\right)^2  \xi_{\phi}	\tag{A4c} \\
\left|\xi_r / \xi_{\phi}\right| &= \frac{\gamma^2 + {k_z}^2 {v_{Az}}^2}{2 \gamma \Omega} \tag{A4d}~~,
\end{align}
where $-i \omega = \gamma \approx k_z v_{Az}$ is the growth rate.

Equation (A4a) is the linearized induction equation with $\mathbf{E}_1 = - \frac{1}{c}\mathbf{v}_1 \times \mathbf{B}$.  
Equations (A4b) and (A4c) are momentum Equations (106) and (107) in BH98, 
obtained by inserting Equation (A4a) into $\frac{1}{c \rho}\left(j_1 \times \mathbf{B}\right) = \frac{1}{4\pi\rho}\left((\nabla\times\mathbf{B}_1) \times \mathbf{B}\right)$. 
Including $B_{\phi}$ gives $(\mathbf{k} \cdot \mathbf{v}_A)^2 = \left[(k_z v_{Az})^2  + (k_{\phi} v_{A\phi})^2\right]$ in Equation (A4b) with no effect in Equation (A4c). 

Combining Equations (A4b) and (A4c) gives the dispersion relation:
\begin{align}
\omega^4 - \omega^2 \left[\kappa^2	+ (k_z v_{Az})^2 + (\mathbf{k} \cdot \mathbf{v}_A)^2\right] &= -\left(k_z v_{Az}\right)^2 [r \frac{\p \Omega^2}{\p r} + (\mathbf{k} \cdot \mathbf{v}_A)^2] \tag{A5a} \\
\kappa^2 &= 4 \Omega^2 + r \frac{\p \Omega^2}{\p r} \tag{A5b}
\end{align}
where $r\frac{\p \Omega^2}{\p r} = - 3{\Omega_K}^2$ for Keplerian rotation. For $B_{\phi} = 0$, 
Equation (A5a) reduces to BH98, Equation (111). Keeping $B_{\phi}$ gives instability with growth rate $\gamma$ if the right hand side of Equation (A5a) is negative, requiring:
\begin{align}
(\mathbf{k} \cdot \mathbf{v}_A)^2 &= (k_z v_{Az})^2  +  (k_{\phi} v_{A \phi})^2 < - r\frac{\p \Omega^2}{\p r}	\tag{A6a} \\
\gamma &= - i\omega	\approx	\left|(k_z v_{Az}) (\mathbf{k} \cdot \mathbf{v}_A)\right|^{1/2} < \Omega_K \tag{A6b}~~,
\end{align}
with the limit on $\gamma$ in BH98. Thus we see that MRI initiated by poloidal flux alone persists even as 
$B_\phi$ grows to exceed $B_z$. Moreover, that adding $B_\phi$ also produces the coupling to perturbations in $z$ 
needed to 
give finite $D_r$ and $D_{\phi}$ in Equation (A2a). We get: 
\begin{align}
-\omega^2 \xi_z &= {k_z}^2 v_{A \phi} v_{Az} \xi_{\phi} \tag{A7a} \\
B_{1 z} &= -i \left[(k_r \xi_r  + k_{\phi} \xi_{\phi}) B_z - k_{\phi}\xi_z B_{\phi}\right] \tag{A7b}
\end{align}
The ratio of omitted electric force terms to magnetic terms can be shown to be 
of order $(v_{\phi}^2/c^2) \ll 1$ at $r > a$.   

Concerned that MRI may not persist, \citet{pariev2007apj} proposed an alternative 
mechanism whereby star-disk collisions provide an external means of sustaining dynamos, 
supported by simulations that maintain a dynamo even when MRI is suppressed \citep{pariev2007}. 
Star-disk collisions could be simulated in GRMHD codes by prescribed, random heat pulses. 

Given any process that can sustain $(-D_{\phi} /D_r) \approx 1$, 
any seed magnetic field around black hole can create an accretion-driven dynamo. 
Whatever the process, it is only necessary that embedded plasma be ``magnetized,'' meaning that the ion 
Larmor radius $r_L$ should be small on spatial coherence scales. For example, if an AGN accretion disk
captures a single magnetized star, like our Sun, it might add a field $0.3$ Gauss, 
giving $r_L = 3 \times 10^7\text{ cm}$ for a maximum gravitational energy of order GeV for hydrogen ions. 
Then a seed field coherent over dimensions $> 3 \times 10^7\text{ cm}$ would grow to astrophysical dimensions. 
Simple estimates show that even advection of primordial plasma can create a seed field, 
with density $\rho_{\text{amb}} \approx 10^{-29}\text{ gm/cm}$ \citep{colgate2004} attached 
by magnetization to the primordial field ($10^{-15}$ gauss, \citet{ando2016}), giving a dynamo 
if the field growth rate $> 1/\tau$, already occurring at a radius $10^{21}\text{ cm}$ about 
equal to $r = R_0$ in Figure 1b, as derived in Paper II, Equation (49). 
This suggests that, in actual disk conditions, MRI itself should be sufficient to sustain the dynamo.

\subsection*{A.4: Extending the Dynamo Inside the Central Column, Cowling's Theorem}

With the approximations above, strong MRI dynamo action exists across the Diffuse Pinch where $\Omega$ is Keplerian and $B_z$ is finite, but the dynamo could die out toward the black hole where the breakdown of our Diffuse Pinch solution indicates a fall off of rotation below Keplerian, even though $B_z(r)$ due to toroidal  current outside $r$ remains large all the way to the black hole. 
That current inside the Central Column is essential to our model of UHECR acceleration is evident from Figure \ref{Fig4},  showing that $I(a)$ is $1/1.7 = 60\%$ of the asymptotic current re-entering the disk at a large radius. 
That the dynamo persists inside the Central Column  is due to the fact that MRI is not local, but should be averaged in radius over the reciprocal $k_r$ not yet specified. 
A plausible value $k_r^{-1} > a$ easily spans the gap from the Diffuse Pinch to the event horizon. 
Moreover, if rotation were to slow down too much for MRI to persist, a term $\frac{\p}{\p t} \int_0^H \, dz 4 \pi \rho_r^2 \Omega$ omitted from Equation (6c) would build up rotation to restore MRI. 
Fortunately, details inside the Central Column are not important for kink stability analysis, as shown in Paper II, and our electric circuit representing the Central Column can include dynamo action at the event horizon \citep{blandford1977,mcdonald1982}.

We conclude that our estimates in Papers I and II are a reasonable guess, giving near the black hole (Equation (65b), Paper II):
\begin{equation}	
E_r(r) \approx \frac{r \Omega}{c}\left|B_z\right| + D_r \approx \frac{r\Omega}{c}B_a  +  D_r \tag{A8}
\end{equation}
and, at the opposite extreme, growing poloidal flux $\Psi$ near the O-point at $r = R_o$, $z = 0$, where the poloidal field vanishes in Figure \ref{Fig1}, given by (Equation (B1), Paper I):
\begin{equation} 	
\frac{\p \Psi(R_0,0)}{\p t} = c R_o D_{\phi}(R_o,0)	\tag{A9}
\end{equation}
To defeat Cowling's anti-dynamo theorem \citep{moffatt1978}, it is shown in Papers I and II that negative $D_{\phi}$ by Equation (A2) becomes positive near $r = R_o$, representing helicity flow from the Central Column to the O-point as in the creation of spheromaks in the laboratory. 

\section*{APPENDIX B: KINETIC INSTABILITY IN AGN JETS; DCLC}

In this Appendix, we review the relativistic derivation of Equation (22) giving, with all terms included, 
two-stream instability in the Central Column and also Drift Cyclotron Loss Cone (DCLC) 
instability in the nose that account for UHECR acceleration in our model.

\subsection*{B.1: Formalism}

Both two-stream and DCLC dispersion relations used in this paper appear in the past literature. 
Here we show their common origin as ion excitation of electron waves, even at relativistic velocities. 
These are electrostatic modes with potential perturbation $\Phi_1$. 
We apply the relativistic Vlasov solution employing Fourier transforms and comment on Landau damping separately. 
The main features emerge in the ``slab'' (Cartesian) coordinate approximation, given by:
\begin{equation}
\Phi_1(\mathbf{x},t) = \Phi_1(x) \, \exp \left\{i \left(k_z z + k_y y  - \omega t \right) \right\} \tag{B1}
\end{equation}
where $z$ is the direction along $\mathbf{B}$, while $y$ lies in the flux surface, and $x$ is perpendicular to a flux surface (the direction of density gradients giving drift modes for DCLC). 

The unperturbed distribution function $F_0$ is given by:
\begin{align}
F_0(P_z, E_{\perp}, P_y) &= n_0 f_0 \tag{B2a} \\
P_z &= p_z \tag{B2b} \\
P_y &= p_y + \left(\frac{q}{c}\right)A_y \tag{B2c} \\
E_{\perp} &= \left[\frac{{p_x}^2 + {p_y}^2}{2 m \gamma_L}\right] = \frac{1}{2} p_{\perp} v_{\perp} \tag{B2d}
\end{align}
In Equation (B2d), $m$ is the rest mass and $P_z$, $E_{\perp}$ and also the total energy $m c^2 \gamma_L$ are relativistic constants of the motion in a uniform magnetic field, while $\gamma_L$ is still constant and the magnetic moment $\propto E_{\perp}/B$ is an adiabatic invariant in a non-uniform field. 
For electrons, dependence of $f_0$ on $P_z$, giving the greatest effect of relativity, yields the two-stream instability, while dependence on $P_y$ adds spatial dependence giving the electron drift waves excited by DCLC. 
The mean-field density $n_0$ is defined by the normalization $\int dP_z \, dE_{\perp} \, dP_y \, f_0 = 1$, and similarly the perturbation $F_1 = n_0 f_1$. 
We first ignore $P_y$ dependence, adequate for ions, giving $\int dP_z \, dE_{\perp} f_0 = 1$. 
Then the perturbation $f_1$ is given by:
\begin{align}
\frac{\p f_1}{\p t} &+ v_y\frac{\p f_1}{\p y} + v_z\frac{\p f_1}{\p z} + q \mathbf{v} \times \mathbf{B} \cdot \frac{\p f_1}{\p \mathbf{p}} \equiv \frac{df_1}{dt} \notag \\
&= -q\left(-\frac{\p \Phi_1}{\p y}\frac{\p f_0}{\p E_{\perp}}\frac{\p E_{\perp}}{\p p_y}\right)  - q\left(-\frac{\p \Phi}{\p z}\frac{\p f_0}{\p P_z}\frac{\p P_z}{\p p_z}\right) \tag{B3a} \\
&= iq \, \Phi_1 \left(k_y \frac{\p f_0}{\p E_{\perp}} v_y  + k_z \frac{\p f_0}{\p P_z}\right) \tag{B3b} \\
\frac{\p E_{\perp}}{\p p_y} &= \frac{1}{2} v_y \left\{ 1 + \frac{1}{{\gamma_L}^2}\left[ 1 + \left(\frac{p_z}{mc}\right)^2\right]\right\} \approx v_y	\tag{B3c}\\
\frac{\p P_z}{\p p_z} &= 1 \quad;\qquad \gamma_L  =  \left[1 + \left(\frac{p}{mc}\right)^2\right]^{1/2} \tag{B3d}
\end{align} 
                                                                                                    
We now integrate Equation (B3a) to obtain $f_1$ and from this the density perturbation $n_1$. 
We follow \citet{fowler1981}, Section VI. 
Several steps are involved. 
First, since $f_1$ is a Fourier transform, we relate $\Phi_1(t')$ at an earlier time $t'$ to $\Phi_1(t)$ by a factor given by:
\begin{align}
&\exp \left\{ i [k_y(y'- y) + k_z(z'- z) - \omega(t'- t)] \right\} \tag{B4a} \\
= &\exp \left\{ i\left[k_y r_L\left(\sin\left[\theta - \omega_c(t'- t)\right] - \sin\theta \right) + (k_z v_z - \omega)(t'- t)\right] \right\} \tag{B4b} \\
\rightarrow  &\langle \sum \exp \left\{ i (k_z v_z -\omega +n\omega_c)(t'- t)\right\} \exp \left\{ i(-n+m)\theta \times J_n(k r_L) J_m(k r_L) \right\} \rangle  \tag{B4c} \\
= &\sum \exp \left\{ i (k_z v_z - \omega + n\omega_c)(t'- t)\right\} Jn(k r_L)^2 \tag{B4d} \\
&\exp \left\{ i k r_L \right\}\sin\theta = \sum J_n(k r_L) \exp \left\{ i n \theta \right\} \tag{B4e}  
\end{align}
Here $\sum$ sums over indices $n$ and $m$ from $-\infty$ to $+\infty$. 
Introducing orbital information into Equation (B4a), following \citep{post1966}, we obtain Equation (B4b), with Larmor spin angle $\theta$ and $z = v_z t$ since $v_z  = (P_z/m\gamma_L)$ and $\frac{d\theta}{dt} = \omega_c = (eB/m\gamma_Lc)$ are constant along the orbit since $P_z$ and the energy $\gamma_L$ are constants of the motion, and we take $B$ to be approximately uniform over the orbit. 
Introducing the identity Equation (B4e) into Equation (B4b) gives Equation (B4c), where $\langle ... \rangle$ anticipates the $\int d\theta$ average in $n_1 = n_0(\textbf{x}) \int d\textbf{p} \, f_1$, giving $m = n$ in Equation (B4d). 
Changing variables to $E_{\perp}$ and $P_z$ and completing this integration gives the perturbed density, with normalization $\int dP_z \, dE_{\perp} \, f_0 = 1$:
\begin{align}
n_1 &= n_0 \int dP_z \, dE_{\perp} \, f_1 \tag{B5a} \\
&= n_0 \int dP_z \, dE_{\perp} \, \int_{-\infty}^t dt' \, iq \Phi_1 \left(k_y \frac{\p f_0}{\p E_{\perp}} v_y' +  k_z \frac{\p f_0}{\p P_z}\right)  \exp \left\{ i [k_y(y'- y) + k_z(z'- z)- 
\omega(t'- t)] \right\} \tag{B5b} \\
&= n_0 \int dP_z \, dE_{\perp} \, q \Phi_1 \int_{-\infty}^t dt' \, \bigg( \frac{\p f_0}{\p E_{\perp}} \left[\frac{d}{dt'} + i(\omega - k_z v_z)\right] \notag \\
& \phantom{{}xxxxxxxxxxxxxxxxxxxxxxx}  + ik_z \frac{\p f_0}{\p P_z}\bigg) \exp \left\{ i [k_y(y'- y) + k_z(z'- z)- \omega(t'- t)]\right\} \tag{B5c} \\
&=	n_0 q\Phi_1 \int dP_z \, dE_{\perp} \, \left\{ \frac{\p f_0}{\p E_{\perp}} + i \left(\omega \frac{\p f_0}{\p E_{\perp}} + k_z \frac{\p f_0}{\p P_z}\right) \int_{-\infty}^t	 dt'\, \sum \left[ 1 - \frac{\omega-k_z v_z }{\omega - k_z v_z - n\omega_c} J_n(k_y r_L)^2 \right] \right\} \tag{B5d} \\
&=	n_0 q\Phi_1 \int dP_z \, dE_{\perp} \, \bigg\{ \frac{\p f_0}{\p E_{\perp}} \left( 1 - \sum \left[\frac{\omega-k_z v_z}{\omega - k_z v_z - n\omega_c} J_n(k_y r_L)^2\right]\right) \notag \\
&\phantom{{}xxxxxxxxxxxxxxxxxxxxxxxxxxxxxxxxx} - \sum k_z \frac{\p f_0}{\p P_z} \left(\frac{1}{\omega - k_z v_z - n\omega_c}\right) J_n(k_y r_L)^2 \bigg\} \tag{B5e} \\
\frac{d}{dt'} & \exp \left\{ i[k_y(y'- y) - (\omega - k_z v_z)(t'- t)] \right\} \notag \\
&= i \left[k_y \frac{dy'}{dt'} - i(\omega - k_z v_z)\right] \exp \left\{ i [k_y(y'- y) - (\omega - k_z v_z)(t'- t)]\right\} \tag{B5f}
\end{align}
Here $'$ denotes quantities at time $t'$. 
The spin angle $\theta$ integration has already been performed in Equation (B5a) giving Equation (B5b), using Equation (B4d). 
Applying Equation (B5f) to replace $ik_y \frac{dy'}{dt'} = i k_y v_y' = \left(\frac{d}{dt'} + i(\omega - k_z v_z)\right)$ gives this operation in the first line of Equation (B5c). 
Carrying out this operation and integrating on $t'$ gives Equation (B5d), in which the leading term no longer involves an orbit integration \citep{post1966}. 
Carrying out the remaining integrations on $t'$ yields the familiar resonance denominators in Equation (B5e). 
Adding dependence on $P_y = (p_y - e/c Ay)$ gives electron drift waves, yielding an additional term of the form:
\begin{align}
\frac{\p(n_0 f_{e0})}{\p P_y} &= n_0 \left(\frac{1}{\p P_y/\p p_y}\right)(\p f_{e0}/\p P_y) + f_{e0} \left(\frac{1}{\p P_y/\p x}\right)(\p n_0 f_{e0}/\p x) \notag \\
& =	n_0 \left(\frac{\p f_{e0}}{\p P_y} +  f_{e0}\frac{\varepsilon c}{qB}\right) \tag{B6}
\end{align}
where $\varepsilon = (\p n_0/\p x)/n_0$. 
Integrating first over the orbit and then over momentum gives the drift contribution to the density perturbation. 
Integrating over momentum eliminates the first term. 
We obtain for the charge density:
\begin{align}
q(n_1)_{\text{drift}} &= \left[\Phi_1 n_0 e^2 \left(\frac{\varepsilon c}{qB}\right)\right] \int dP_z \, dE_{\perp} \, \int_{-\infty}^t dt' \, f_{e0} i k_y \exp \left\{ i [k_y(y'- y) + k_z(z'- z) - \omega(t'- t)] \right\} \tag{B7a} \\
&= \Phi_1 n_0 e^2 k_y\left(\frac{\varepsilon c}{qB}\right) \sum \frac{J_n(k_y r_L)^2}{\omega - k_z v_z - n\omega_c} \tag{B7b} \\
&\rightarrow \Phi_1 n_0 e^2 k_y\left(\frac{\varepsilon c}{qB \omega}\right)	= \frac{\Phi_1  k_y \varepsilon}{4\pi} \left(\frac{{\omega_{pe}}^2}{\omega_{ce}\,\omega}\right) \tag{B7c}
\end{align}
Equation (B7c) gives the drift term to be added to Equation (B5b). 
Integration on $t'$ gives Equation (B7b) by analogy with Equation (B5c) (but missing the operation $d/dt'$ coming from a factor $v_y'$ in that equation).  
Taking $k_y r_{L e} \ll 1$ for electrons gives Equation (B7c), rewritten on the far right hand side as it usually appears in the literature \citep{post1966}, though in fact this term does not contain mass so that there is no relativistic correction. 
Introducing (B7c) into Poisson's Equation, $- \nabla^2\Phi_1 = 4\pi (e n_{1i} - e n_{1e})$, and dividing by $k_y^2$ gives the right hand side of Equation (22b) giving the electron drift wave contribution in the DCLC dispersion relation in Section 3.3.2. 
The electron term with $\frac{\p f_0}{\p P_z}$ in Equation (B5e) will give two-stream instability in Appendix C.

\subsection*{B.2: Relativistic DCLC Instability in the Nose}
	
Growth of $v_y$ or a pre-existing $v_y$ adds ion cyclotron resonance terms to the left hand side of Equation (22a). 
Resonance gives large contributions from the ion Bessel functions in Equation (B5), but only if $k_y r = 1$ inside the plasma. 
That this cannot happen in the Central Column follows from:
\begin{equation}
k_y a = \varepsilon a \left(\frac{\omega_{pe0}^2}{\omega_{ce0} \omega_{ci0}}\right) \gamma_{CC}    = 4.4 \times 10^{-4} M_8 \ll 1 \tag{B8}
\end{equation}
where the subscript $0$ denotes rest mass quantities. 
We see that $k_y a = 1$ is not possible for any known black hole mass. 
In laboratory experiments, these ion cyclotron modes occurred when the non-relativistic $\omega_{ci} < \omega_{pi}$ \citep{post1981}; or equivalently, when: 
\begin{equation}
\left(\frac{\omega_{pi}}{\omega_{ci}}\right)^2 = \beta_{i \perp}\left(\frac{c}{v_{\perp}}\right)^2 > 1 \tag{B9}
\end{equation}
Equation (B9) is an approximate condition for DCLC instability with pressure parameter $\beta_{i \perp}$ in Equation (20a). 
This can be satisfied in the laboratory at low $\beta_{i \perp}$ if $v_{\perp} \ll c$, but in AGN jets it is satisfied only as ions enter the nose. 

A DCLC dispersion relation like the non-relativistic form giving the DCLC instability threshold in Equation (23a) can be derived from Equation (B5e) applied to ions. 
The non-relativistic derivation, in \citet{post1966} reproduced in \citet{fowler1981}, follows the same steps that led to Equation (B5e). 
Relativity would appear only in the mass defining the cyclotron frequency. 
The result, using also the drift term discussed above, is given by: 
\begin{align}
k_y^2 &= k_y \varepsilon \left(\frac{{\omega_{pe0}}^2}{\omega_{ce0} \, \omega}\right) \tag{B10a} \\
&- {\omega_{pi0}}^2  \int dP_z \, dE_{\perp} \, \left\{ m_i \frac{\p f_0}{\p E_{\perp}} \left[1- \sum J_n(k_y r_{L i})^2 \frac{\omega}{\omega - n\omega_{ci}}\right] \right\} \tag{B10b}
\end{align}
Instability requires $\frac{\p f_0}{\p E_{\perp}} > 0$ near $E_{\perp} = 0$ representing the ``hole'' in perpendicular energy discussed in Section 3.3.2. 
It was found that, mathematically, a sufficient hole is present when the large Larmor orbits cause the $P_z$-averaged distribution $\int dP_z \, dE_{\perp} \, \frac{\p f_0}{\p E_{\perp}}$ to be strongly positive while a Maxwellian in $E_{\perp}$ would give a negative value. 
Given a hole, Equation (B10a, b) is identical with the non-relativistic formulation, aside from swapping $P_z$ for $v_z$ in the integration, giving relativistic masses but otherwise identical with the DCLC instability in non-relativistic form with relativistic masses, as in Equation (23a). 

\section*{APPENDIX C: ELECTRON SYNCHROTRON RADIATION, TWO-STREAM INSTABILITY}

In this Appendix, we discuss synchrotron radiation, beginning with the dispersion relation for two-stream instability that converts MHD kink mode acceleration of electrons in the Central Column into orbital spin giving rise to synchrotron radiation, and concluding with the synchrotron energy spectrum.

\subsection*{C.1: Two-Stream Instability}

To obtain the dispersion relation giving two-stream instability between counter-streaming ions and electrons in the Central Column, we set $\varepsilon = 0$ in Equation (B6) giving $n_1$ by Equation (B5e) with $k_y =  0$ and we take the limit $k_y r_L \rightarrow 0$ in all Bessel functions giving $J_o \rightarrow 1$ and $J_n \rightarrow 0$ for all $n \neq 0$. 
We assume ions are accelerated to velocity $v_z = c$ and electrons to velocity $v_z = - c$. 
Poisson's equation becomes:
\begin{align}
{k_z}^2 \Phi_1 &= 4 \pi (q_e n_{1e}  + q_i n_{1i}) \tag{C1a} \\
	&= - 4\pi n_0 e^2 \Phi_1 k_z \int dP_z \, dE_{\perp} \, \sum_{j=e,i}\frac{\p f_{0j}}{\p P_z}(\omega - k_z v_z)^{-1} \tag{C1b} \\
1 &= \left(\frac{{\omega_{pe0}}^2}{{\gamma_{Le}}^3}\right) f_{e0} \frac{1}{(\omega + k_z c)^2} + \left(\frac{{\omega_{pi0}}^2}{{\gamma_{Li}}^3}\right) f_{0i} \frac{1}{(\omega - k_z c)^2} \tag{C1c}
\end{align}
In Equation (C1b), setting $k_y =  n = 0$ has eliminated terms with $\frac{\p f_0}{\p E}$ giving $n_1$ in Equation (B5e). 
Equation (C1c) follows on dividing by ${k_z}^2 \Phi_1$ after applying the following result in \citet{montgomery1964}, Equations (10.51) and (10.52). 
Using $P_z = p_z$ and $\gamma_L$ from Equation (B3d) with relativistic $v_z = (p_z/m\gamma_L)$, we obtain:
\begin{align}
\frac{\p f_0}{\p p_z}\left(\frac{1}{\omega - k_z v_z}\right) &= \frac{\p}{\p p_z}\left(f_0 \frac{1}{\omega - k_z v_z}\right) -  f_0 \frac{\p}{\p p_z}\left(\frac{1}{\omega - k_z v_z}\right) \tag{C2a} \\
&\rightarrow  \frac{-f_0}{\left(\omega - k_z v_z\right)^2} \frac{\p}{\p p_z}\left(k_z v_z\right) \tag{C2b} \\
m \frac{\p v_z}{\p p_z} &= \frac{1}{{\gamma_L}^3}\left[{\gamma_L}^2 - (p_z^2/m^2c^2)\right] = \frac{1}{{\gamma_L}^3}\left[1 + (p_{\perp}^2/m^2c^2)\right] \approx \frac{1}{{\gamma_L}^3} \tag{C2c}
\end{align}
where Equation (C2b) is the surviving term in an integration by parts.

Combining Equations (C2b) and (C2c) gives the final two-stream dispersion relation, Equation (C1c), with the assumption that there is no acceleration perpendicular to $\textbf{B}$, giving the far right hand side of Equation (C2c).

We note that the ion term in Equation (C1c) is largest at resonance, suggesting that we take $\omega = k_z c + \Delta \omega$ near the ion resonance (Goldston \& Rutherford 1995, Section 23.4 ). 
Upon substituting $\omega = k_z c + \Delta\omega$ and expanding, Equation (C1c) gives: 
\begin{align}
{k_z}^2 &= {k_z}^2 \left[\frac{{\omega_{pe}^*}^2}{(\omega + k_z c)^2} + \frac{{\omega_{pi}^*}^2}{(\omega - k_z c)^2}\right] \tag{C3a} \\
&= {k_z}^2 \frac{{\omega_{pe}^*}^2}{(2k_z c)^2}\left( 1 - \frac{\Delta \omega}{k_z c} + ... \, \right) + \frac{{\omega_{pi}^*}^2}{\Delta \omega^2} {k_z}^2 \tag{C3b} \\
k_z c &= \frac{1}{ 2}  \, \omega_{pe}^* \tag{C3c} \\
\Delta \omega^3	&= (k_z c \,{\omega_{pi}^*}^2)	=  (k_z c)^3 \frac{{\omega_{pi}^*}^2}{2{\omega_{pe}^*}^2}  =  (k_z c)^3\frac{8 m_e}{m_i} \tag{C3d} \\
\omega &= k_z c + \Delta\omega = k_z c \left[ 1 +  \left(\frac{8 m_e}{ m_i}\right)^{1/3}(\cos 60^{\circ} + i \sin 60^{\circ})\right] \tag{C3e}
\end{align}
where $\omega$ in Equation (C3e) is the unstable root. 
Here ${\omega_{p}^*}^2 = (4 \pi n_0 e^2 / m {\gamma_L}^3)$. 
In the Central Column, radiation-limited acceleration gives $\gamma_{Li} \geq \gamma_{Le}$ by Equation (18) including electron synchrotron radiation, giving then the ratio of rest masses in Equations (C3d) and (C3e) so that the only relativistic effect is the speed of light as the wave velocity.

\subsection*{C.2: Electron Scattering} 

Once initiated, the two-stream instability can scatter electrons to transfer parallel acceleration to synchrotron radiation, as in Section 5.1. 
To do so, the electron beam must excite waves with non-zero $k_y$ which adds terms to the dispersion relation in Equation (C1c). 
Keeping both the electron drift term from Equation (B7c) and potentially-resonant ion terms with finite $k_y$, we obtain: 
\begin{align}
{k_y}^2 + {k_z}^2 &= \left\{\int dP_z \, dE_{\perp} \, f_{e0} \left[\frac{{k_z}^2}{{\gamma_{Le}}^3}\frac{{\omega_{pe0}}^2}{(\omega - k_z v_z)^2} + k_y \varepsilon \frac{{\omega_{pe0}}^2}{\omega_{ce0}\omega}\right] \right\}_{\text{wave}} \tag{C4a} \\
&+ \left\{ L_L + \left[{k_z}^2 \frac{\left({\omega_{pi0}}^2/{\gamma_{CC}}^3\right)}{(\omega - k_z c)^2}\right] + {k_y}^2 \left[J_1(k_y r_{L i})^2 \frac{{\omega_{pi0}}^2}{(\omega - k_z c - \omega_{c i})^2}\right]\right\}_{\text{drive}} \tag{C4b} \\
L_L &= i\pi \left({\omega_{pe0}}^2 {m_e}^2 \right) \left\{ \int dP_z \, dE_{\perp} \, \delta(P_z - \frac{m_e \gamma_{Le} \omega}{k_z}) \gamma_{Le} \frac{\p f_{e0}}{\p P_z}\right\} \tag{C4c} \\ 
&\approx i\pi \left(\frac{{\omega_{pe0}}^2}{c^2}\right) \left[(m_e^2 c^2 )\frac{\p f_{e0}}{\p P_z}\right]_{\text{res}} \tag{C4d}
\end{align}
Terms labeled ``wave'' generate the waves driven unstable by the terms labeled ``drive,'' with $\gamma_{CC}$ from Equation (19b). 
In the absence of scattering, $f_{e0}$ becomes a delta-function giving the result  in Equation (C4a). 
Ions do not scatter, as discussed in Section 3.2, giving in the drive term in Equation (C4b) the same ion term as that in Equation (C2a), together with a new term taking into account cyclotron resonance. 
This ion cyclotron term comes from Equation (B5e) but dropping the $n = 0$ term of order $k_y^2(k_y^2 r_{L i}^2)$ compared to $k_y^2$ on the left hand side of Equation (C4a). 

Lastly, we include in the drive term an imaginary contribution $L_L$ from the electron Landau pole, whereas the integral in Equation (C4a) is to be interpreted as the principal part. 
The factor $\gamma_{Le}$ in the Landau term arises from changing variables from $v_z$ to $P_z$. 
The Landau pole gives damping or inverse damping depending on signs. 
The magnitude of the Landau damping term in Equation (C4d), comparable to other electron terms and much larger than ion terms away from resonance, indicates that Landau damping would cause electron momentum diffusion to stop short of the ion resonance. 
Resonance extends between positive $1 > v_z/c > (1 - \Delta \omega/k_zc)$, giving, by Equation (3e), the following allowed spread of electron velocities with no Landau damping: 
\begin{align}
1 >	\frac{v_{ze}}{c} > 	\left(1 - \frac{\Delta\omega}{k_z}\right) =  \left[ 1 - \left(\frac{m_e}{2m_i}\right)^{1/2}\cos 60^{\circ}\right] = 0.99 \tag{C5}
\end{align}
Thus, as noted in Section 3.2, a uniform spread in electron momentum gives negligible net velocity and negligible electron current while avoiding electron resonance giving Landau damping. 
That Landau damping exists but ceases if there are few resonant electrons has been demonstrated conclusively in laboratory experiments in the low density regime relevant to AGN jets \citep{post1981}. 

A necessary condition is that $k_y$ (surrogate for $k_r$ in cylindrical coordinates) 
fit inside the Central Column. 
Keeping only the wave terms in Equation (C4a), the dispersion relation in Equation (22) reduces to:
\begin{align}
k^2 &= {k_y}^2 + {k_z}^2 \approx {k_z}^2 \left(\frac{{\omega_{pe0}}^2}{\omega^2}\right) F + k_y \varepsilon \left(\frac{{\omega_{pe0}}^2}{\omega_{ce0}\,\omega}\right) \tag{C6a} \\
\frac{10}{{\gamma_{CC}}^3} &< \quad F \approx \int_1^{\gamma_{CC}} d\gamma_L \frac{1}{\Delta\gamma_L}\left(\frac{{\gamma_{L \perp e}}^2}{{\gamma_{Le}}^3}\right) \quad < \frac{10}{\gamma_{CC}} \tag{C6b} \\
\omega_{pe0} =	2.3 \times 10^3 \, {M_8}^{5/8} \quad&;\qquad \omega_{ce0} =  2.7 \times 10^{10}{M_8}^{-1/2} \quad;\qquad  \omega_{ci} =  \frac{\omega_{ci0}}{\gamma_{CC}} = 0.4 {M_8}^{1/8}  \tag{C6c} \\
n_0 = \frac{I}{\pi a^2 e c} &= 1.7 \times 10^{-3} {M_8}^{-3/2} \quad;\qquad	\gamma_{CC} = 3.4 \times 10^7 {M_8}^{5/8} \tag{C6d}
\end{align} 
The quantity $F$ in Equation (C6b) approximates $f_0$ in the principal-part integration in Equation (C6a) as a constant over a momentum spread $\propto \Delta\gamma_L \approx \gamma_{CC}$ limited by curvature radiation, in Equation (19b). 
The strong weighting at small $\gamma_L$ gives the result shown 
with $\gamma_{L\perp e} = 3$. 
Examining these results shows that waves fit with $k_y = k_z$ and $k_y a = 1$. 

\subsection*{C.3: Electron Transport in the Central Column}
 
Synchrotron radiation mainly arises due to acceleration of electrons by MHD kink modes in the Central Column, as discussed in Section 3.1.  
Electron acceleration is equal and opposite to ion acceleration in Section 3.2. 
The counter-streaming electrons and ions excite the two-stream instabilities that scatter electrons to produce also electron synchrotron radiation. 
We describe these electron processes by a quasi-linear transport equation analogous to Equation (29b) for ions in the nose, now omitting spatial derivatives but keeping $p_{\parallel}$  ($= p_z$ above) as an approximate constant of the motion and transforming $P_{\mu}$ into $p_{\perp}$. 
We obtain:
\begin{align}\nonumber
\frac{\p f_{e0}}{\p t} &+ \left\{\frac{\p }{\p p_{\parallel}}[(q(\Delta V/L)-qE_\mathrm{curv})f_{e0}-D_{p||}\frac{\p f_{e0}}{\p p_{\parallel}}]+T\right\}_{||}
\\ &- \left\{T+\frac{\p}{\p p_{\perp}}[qE_\mathrm{syn}f_{e0}-D_{p_\perp}\frac{\p f_{e0}}{\p p_{\perp}}] \right\}_\perp = 0 \hspace{0cm} 
\tag{C7a} \\
T&=-\frac{\p }{\p p_{\parallel}} \left[D_\mathrm{scat}\frac{\p f_{e0}}{\p p_{\parallel}} \right]
\tag{C7b} \\
E_{\text{syn}} &= \frac{2}{3} e \left(\frac{{\gamma_{Le}}^4}{{r_{Le}}^2}\right)\left(\frac{v_{\perp}}{c}\right)^4 = \frac{2}{3} e \left[\frac{{\gamma_L}^2}{{r_{L1}}^2}\left(\frac{v_{\perp}}{c}\right)^2\right]_e	= \frac{2}{3} e {\left(\frac{p_{\perp}}{m c r_{L1}}\right)_e}^2 \tag{C7c}
\end{align}
where $D_{p \parallel}$ determines the energy spread by two-stream instability limited by curvature radiation $E_\mathrm{curv}$ giving Equation (19b) and  $D_\mathrm{scat}$ describes scattering from $p_{||}$ to $p_\perp$ that yields synchrotron radiation. In Equation (C7c), $E_\mathrm{syn}$ is electron synchrotron 
%
radiation (Equation (18), $\beta_L = v_{\perp}/c$) with electron Larmor radius $r_{L e} = (p_{\perp}c/e B_a)$ and $r_{L1} = (m_e c^2/e B_a$) is constant.

Qualitatively, assuming factorability $f_{e0}=f_{e0||}f_{e0\perp}$ gives $f_{e0||}$ by setting the bracket $\{...\}_{||}=0$ and $f_{e0\perp}$ by setting the bracket $\{...\}_{\perp}=0$, with coupling via $T$. Strong coupling gives $f_{e0||}$ relatively flat in the domain in Equation (C5), with a tail terminating at TeV's due to $E\mathrm{curv}$; and $f_{e0\perp}= (1/eE_\mathrm{syn})\int_0^{p_\perp}d p_{\perp}'T $ perhaps best approximated by a power law as in \citet{li2000}. By contrast, dominant $D_{p\perp}$ with $E_\mathrm{syn}\propto p_\perp^2$ yields $f_{e0\perp}= C \exp [- (p_\perp/p_0)^3] $ with $p_0$ determined by the transport coefficients.  

\end{document}